\documentclass[twoside,twocolumn,9pt]{article}
\usepackage{extsizes}
\usepackage[super,sort&compress,comma]{natbib}
\usepackage[left=1.5cm,right=1.5cm,top=1.8cm,bottom=2.0cm]{geometry}
\usepackage{times,mathptmx}
\usepackage{microtype}
\usepackage{amsmath,amssymb,bm}
\usepackage{graphicx}
\usepackage{xcolor}
\usepackage{sectsty}
\usepackage[format=plain,justification=raggedright,singlelinecheck=false,
            font=small,labelfont=bf,labelsep=space]{caption}
\usepackage[hidelinks]{hyperref}
\usepackage[T1]{fontenc}

\providecommand{\doi}[1]{}
\renewcommand{\doi}[1]{%
  \href{https://doi.org/#1}{doi:\,\nolinkurl{#1}}%
}

\sectionfont{\large\bfseries}
\subsectionfont{\normalsize\bfseries}
\newcommand{\jump}[1]{[\![#1]\!]}
\newcommand{\ReE}{\mathrm{Re}_{E}}
\newcommand{\CaE}{\mathrm{Ca}_{E}}
\newcommand{\El}{\mathrm{El}}
\newcommand{\Ma}{\mathrm{Ma}}
\newcommand{\Dq}{D_q}
\newcommand{\DG}{D_{\Gamma}}
\newcommand{\Ks}{K_s}
\newcommand{\bI}{\bm{I}}
\newcommand{\bn}{\bm{n}}
\newcommand{\bu}{\bm{u}}
\newcommand{\bE}{\bm{E}}
\newcommand{\ber}{\bm{e}_r}
\newcommand{\bet}{\bm{e}_{\theta}}
\newcommand{\grad}{\bm{\nabla}}

\begin{document}

\twocolumn[
\begin{@twocolumnfalse}
\noindent\LARGE\textbf{Immobilisation and remobilisation of surfactant-covered drop interfaces in electric fields: Simulations}

\vspace{0.55cm}
\noindent\large\textbf{Debasish Das$^{\ast a}$}

\vspace{0.25cm}
\noindent\small\textit{$^{a}$ Department of Mathematics and Statistics, University of Strathclyde, Livingstone Tower, 26 Richmond Street, Glasgow G1 1XH, United Kingdom.}

\noindent\small$^{\ast}$\textit{E-mail: debasish.das@strath.ac.uk}

\vspace{0.35cm}
\noindent\textbf{Abstract}\par
\vspace{0.08cm}
\noindent\small
Fluid--fluid interfaces are seldom clean due to the presence of surfactants, which can strongly modify their dynamics. Interfacial flow redistributes surfactant, creating concentration gradients whose Marangoni stresses can immobilise the interface. By contrast, surface diffusion or exchange with the bulk can reduce these gradients and restore interfacial motion. We study this competition using a two-dimensional model of a surfactant-covered leaky-dielectric drop in a uniform electric field. Electric stresses drive the classical Taylor circulation, in which the drop remains non-rotating. At stronger fields, convection of interfacial charge can destabilise this state and produce the Quincke regime, characterised by spontaneous mean rotation. Using a fully nonlinear model on a fixed circular interface, we show that weak surfactant diffusion allows Marangoni stresses to suppress the Taylor circulation almost completely. Quincke rotation remobilises this nearly arrested interface and restores a substantial straining circulation, whose orientation varies with the applied field. Surfactant also shifts the critical field for rotation and can change the onset from subcritical to supercritical. Surface diffusion and exchange with a well-mixed bulk reservoir can produce similar flow responses and rotation thresholds despite different surfactant distributions. The spatial variation of surfactant concentration therefore provides additional information for distinguishing the two transport mechanisms.
\vspace{0.55cm}
\end{@twocolumnfalse}
]

\section{Introduction}

Surfactants are common in industrial and natural multiphase systems, where they are used to stabilise emulsions and foams, control coating and wetting processes, and regulate the motion and coalescence of drops and bubbles.\citep{Langevin2026,ManikantanSquires2020} Even when present at low concentrations, surfactants can strongly affect the flow because they preferentially adsorb at fluid interfaces. Spatial variations in surfactant concentration produce variations in surface tension and hence Marangoni stresses, which modify the motion of the interface and the surrounding fluids.\citep{rosen2012surfactants}

A well-known consequence of this coupling is the immobilisation of a fluid interface by Marangoni stresses, as described in the classical theory of \citet{FrumkinLevich1947}. Tangential flow redistributes adsorbed surfactant and can cause it to accumulate near regions where the surface flow converges. The resulting surface-tension gradients oppose this motion, reducing the surface velocity and, when sufficiently strong, effectively immobilising the interface. The degree of immobilisation depends on both surfactant redistribution and the sensitivity of surface tension to concentration. Combining theory and experiments on rising surfactant-laden bubbles, \citet{Palaparthi2006} examined the stagnant-cap regime, in which surfactant accumulation immobilises the rear of the bubble while the front end of the surface remains mobile. Their work identifies how surfactant transport and exchange control the extent of the stagnant region. For an insoluble surfactant on a drop in an extensional flow, \citet{Eggleton1999} found stagnant surfactant-rich regions adjoining mobile, depleted regions at low coverage. At high coverage, the steep dependence of surface tension on concentration allowed strong Marangoni stresses and complete stagnation at steady state despite relatively small concentration variations. For flow over superhydrophobic surfaces, \citet{MayerCrowdy2022} likewise distinguished stagnant-cap formation from immobilisation associated with the approach to maximum surfactant packing and the resulting steep nonlinear equation of state.

Surfactant transport also influences drop deformation and migration. \citet{StoneLeal1990} examined the competition between surfactant convection and dilution caused by the increase in interfacial area during deformation. For soluble surfactants, \citet{ChenStebe1997} showed how the adsorption framework and exchange kinetics affect the retardation of thermocapillary migration. \citet{EggletonStebe1998} studied a deforming drop in an extensional flow and showed that adsorption and desorption can alter both the Marangoni stresses and the distribution of surface tension, producing a non-monotonic dependence of deformation on the mass-transfer rate at high coverage.

Interfacial mobility can be restored when transport relaxes the concentration gradients generated by the flow. \citet{Stebe1991} demonstrated remobilisation for surfactants with rapid adsorption--desorption kinetics and sufficiently effective bulk transport. In their experiments, micellar solutions of Triton X-100 and Brij-35 recovered mobile interfaces at high concentrations, whereas the slowly desorbing protein bovine serum albumin did not. \citet{StebeMaldarelli1994} extended this approach to mixtures containing a retarding surface-active component. By adding a rapidly exchanging surfactant in sufficient quantity, competitive adsorption allowed it to dominate the interface and restore mobility. Surface diffusion provides another means of relaxing concentration gradients. The possibility that different combinations of surfactant properties and transport processes can produce similar observable flows underlies the ``hidden-variable'' perspective of \citet{ManikantanSquires2020}. Measurements of interfacial motion alone may therefore be insufficient to identify the mechanisms controlling that motion.

Electric fields drive interfacial flow directly, providing a setting in which to study how surfactants modify circulation and interfacial mobility. In the Taylor--Melcher model of a leaky-dielectric drop, the bulk liquids are assumed to be electroneutral, so free charge is confined to the liquid--liquid interface and is supplied by weak but finite Ohmic currents in the two phases.\citep{Taylor1966,MelcherTaylor1969,Saville1997} The electric field acting on this interfacial charge produces a tangential Maxwell stress that drives Taylor circulation inside and outside the drop. Depending on the electrical properties of the two liquids, the surface flow is directed either from the equator towards the poles or from the poles towards the equator. This circulation defines the non-rotating Taylor regime; normal electric stresses additionally deform a freely suspended drop. In the complete leaky-dielectric model, interfacial charge conservation balances the local rate of change of surface-charge density and surface-charge convection against the jump in normal Ohmic current. Charge convection can produce sharp charge-density gradients in the Taylor regime.\citep{Lanauze2015,DasSaintillan2017small,DasSaintillan2017,Peng2024} A surfactant-covered leaky-dielectric drop subject to a uniform electric field therefore contains two coupled interfacial fields: surface charge and surfactant concentration. The same interfacial velocity redistributes both quantities, while electric and Marangoni stresses feed back on that velocity. This creates a strongly nonlinear coupling between the electrical, hydrodynamic and surfactant dynamics.

The electrohydrodynamics of surfactant-covered drops has been studied mainly in regimes where surface-charge convection is neglected or treated as a weak perturbation. \citet{HaYang1995} analysed the deformation and stability of a drop covered by insoluble surfactant in an applied electric field. \citet{TeigenMunkejord2010} subsequently used numerical simulations to show that surfactant can either increase or decrease drop deformation depending on the direction of the Taylor circulation. \citet{Nganguia2013} developed small-deformation and spheroidal models for the equilibrium electrodeformation of a surfactant-covered drop. In their non-diffusing equilibrium limit, the interface becomes completely immobilised: Marangoni stress balances the tangential electric stress and suppresses the Taylor circulation. \citet{Nganguia2019} studied nonlinear surfactant advection and diffusion in electrodeformation, while \citet{Nganguia2021} included adsorption, desorption and bulk surfactant transport; both studies neglected surface-charge convection. Three-dimensional boundary-integral simulations have also treated general drop shapes and interacting drops in surfactant electrohydrodynamics.\citep{Sorgentone2019,SorgentoneVlahovska2021} Charge convection has been retained together with surfactant transport in perturbative calculations; for example, \citet{Poddar2018} considered a sedimenting drop using a small electric Reynolds-number expansion. These studies establish the influence of surfactant transport on electrically driven deformation and circulation, primarily under conditions of negligible or weak charge convection. Interfacial rheology also modifies electrically driven drop dynamics. \citet{Nganguia2023} used a spheroidal model to show how surface shear and dilatational viscosities alter prolate drop deformation and circulation, with their effects depending on the electrical properties and viscosity ratio.

At stronger electric fields, surface-charge convection can destabilise the Taylor state and produce spontaneous Quincke rotation.\citep{quincke1896} In the Quincke regime, the drop has a nonzero mean rotational motion, with a tilted electric dipole and an electric torque that balances viscous resistance. Straining circulation can coexist with this mean rotation. When charge relaxes more slowly inside the drop than in the suspending liquid, the induced dipole can oppose the applied field. A small tilt then creates an electric torque, and charge convection can reinforce the tilt until sustained rotation develops. Quincke rotation also provides a versatile platform for synthetic active matter. Near a solid boundary, rotating solid spheres translate as Quincke rollers, whose collective dynamics include directed motion\citep{Bricard2013} and coherent vortices.\citep{Bricard2015} Subsequent developments involving Quincke rollers include run-and-tumble motion,\citep{Karani2019} chiral pattern formation,\citep{ZhangSokolovSnezhko2020} colloidal oscillations,\citep{Zhang2021QuinckeOscillations} dynamic roller clusters,\citep{MauleonAmieva2023} magnetic control,\citep{ReyesGarza2023} and anisotropic rolling.\citep{Fitzgerald2025} Geometrical asymmetry can also convert Quincke rotation into propulsion in a bulk fluid.\citep{DasLauga2019}

Quincke rotation is also observed in liquid drops and experiments have reported steady rotation, tilted shapes and, for sufficiently low drop viscosities, hysteretic onset.\citep{KrauseChandratreya1998,HaYang2000,SalipanteVlahovska2010,SalipanteVlahovska2013} Fully nonlinear three-dimensional simulations of the complete leaky-dielectric model reproduced the transition from Taylor circulation to Quincke rotation and demonstrated the central role of charge convection.\citep{DasSaintillan2017} The three-dimensional small-deformation model of \citet{HeSalipanteVlahovska2013} retained charge convection associated with the rotational part of the interfacial velocity but neglected convection by the straining flow. The later theory of \citet{DasSaintillan2021} retained the leading-order contribution of the straining flow to surface-charge convection. More recently, \citet{McDougall2026} extended this framework to an insoluble surfactant-covered drop by coupling charge convection to surfactant transport and Marangoni stresses. Their near-spherical model retains the first few interfacial harmonics, applies when the straining-flow contribution to charge convection remains weak, and predicts that surfactant can shift the critical field for Quincke rotation.

To our knowledge, a corresponding fully resolved three-dimensional calculation coupling nonlinear charge convection and surfactant transport in the low-viscosity-ratio Quincke regime has not yet been reported. Such a calculation would require tracking a deforming interface while resolving sharp surface-charge gradients and strongly non-uniform surfactant distributions. Systematic parameter studies using pseudo-arclength continuation would consequently be computationally demanding. We therefore adopt a two-dimensional model with a fixed circular interface, motivated by the strong surface-tension regime in which deformation is small, while retaining nonlinear charge convection and surfactant transport. The surface charge and surfactant concentration are resolved around the interface without assuming weak charge convection or restricting the fields to a few harmonics. The circular-drop model of \citet{Feng2002} and the large electric Reynolds-number analysis of \citet{YarivFrankel2016} established that this geometry retains the nonlinear mechanism responsible for electrorotation. \citet{Lanauze2015} and \citet{DasSaintillan2017small} reported the development of sharp surface-charge gradients under strong charge convection. More recently, \citet{Peng2024} analysed charge-density blowup in the circular-drop problem without charge diffusion, identifying the equatorial singularity and the associated balance of charge transport and annihilation. \citet{PengSchnitzer2025} demonstrated that the onset of Quincke rotation at low viscosity ratio can become subcritical, producing a bistable field interval, qualitatively consistent with the experimental hysteresis reported by \citet{SalipanteVlahovska2010}. Despite these advances for clean drops, to our knowledge the fully nonlinear coupling between surface-charge convection and surfactant transport through the Taylor--Quincke transition has not yet been investigated in a spatially resolved calculation even within the fixed-circle model. The circular geometry therefore provides a tractable setting in which to address this problem, resolve the coupled interfacial distributions and follow steady solution branches through the onset of rotation.

We investigate how surfactant modifies these dynamics when charge convection is appreciable and Quincke rotation is possible. We examine the suppression of Taylor circulation by Marangoni stresses, the recovery of interfacial motion and straining circulation when rotation begins at weak surfactant diffusion, and the effects of surfactant on the critical field and the nature of the rotational bifurcation. Motivated by the hidden-variable perspective of \citet{ManikantanSquires2020}, we then compare surface diffusion with exchange through a well-mixed bulk reservoir to determine whether the two transport mechanisms can produce similar flow responses and whether their concentration distributions distinguish them. The model couples surface-charge conservation, including nonlinear charge convection, to surfactant transport and a Langmuir equation of state. Steady solutions and their linear stability are used to characterise immobilisation, remobilisation and the transition between the Taylor and Quincke regimes. These calculations extend the reduced descriptions discussed above to spatially resolved nonlinear interfacial transport within the fixed-circle model, providing a test of whether similar flow responses can conceal distinct surfactant-transport mechanisms.

\section{Dimensional problem formulation}
\label{sec:dimensional}

All quantities in this section are dimensional. Dimensionless variables are introduced in Sec.~\ref{sec:nondimensional}, where the same symbols are reused for brevity.

\subsection{Geometry and electric problem}

We consider an infinitely long cylindrical drop of radius $a$ suspended in an immiscible, unbounded liquid. The problem is two-dimensional in the plane normal to the cylinder axis. The drop occupies $V^- = \{r<a\}$ and the suspending liquid occupies $V^+ = \{r>a\}$. Superscripts $-$ and $+$ denote fields inside and outside the drop, respectively. Subscripts $r$, $\theta$ and $n$ denote vector or tensor components and do not denote differentiation; derivatives are written explicitly using $\partial/\partial r$, $\partial/\partial\theta$ and $\partial/\partial t$. The two liquids are Newtonian leaky dielectrics with permittivities $\epsilon^\pm$, electrical conductivities $\sigma^\pm$ and dynamic viscosities $\mu^\pm$. 
A uniform direct-current electric field $E_0\bm{e}_z$ is imposed at infinity. Polar coordinates $(r,\theta)$ are centred on the drop, with $\theta=0$ along the positive $z$-direction, so that $z=r\cos\theta$ and $x=r\sin\theta$. The outward unit normal is $\bn=\ber$ and the unit tangent in the direction of increasing $\theta$ is $\bet$. For any field with limiting values on the two sides of the interface we use the jump convention $\jump{f}=f^+-f^-$. We denote the angular average of an interfacial quantity by
\begin{equation}
 \langle f\rangle=\frac{1}{2\pi}\int_0^{2\pi}f(\theta)\,d\theta.
 \label{eq:surface_average}
\end{equation}

\begin{figure}[t]
 \centering
 \includegraphics[width=0.9\columnwidth]{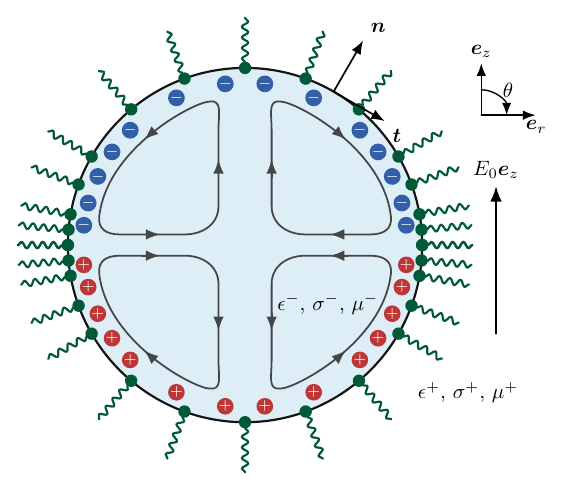}
 \caption{Schematic of the two-dimensional problem. A surfactant-covered circular leaky-dielectric drop of radius $a$ is placed in a uniform electric field $E_0\bm e_z$. Representative surface charges and the internal Taylor circulation are shown for electrical parameters with $RS>1$, for which the interfacial flow is directed from the poles towards the equator. The interface is fixed at $r=a$.}
 \label{fig:schematic}
\end{figure}
The bulk liquids are assumed charge-free, so that the electric fields are irrotational and the electric potentials $\phi^\pm$ satisfy
\begin{equation}
 \bE^\pm=-\grad\phi^\pm,
 \qquad
 \nabla^2\phi^\pm=0
 \quad \text{in }V^\pm.
 \label{eq:dim_laplace}
\end{equation}
The imposed field gives the far-field condition
\begin{equation}
 \phi^+ \longrightarrow - E_0 r\cos\theta, \qquad r \longrightarrow \infty.
 \label{eq:dim_phi_far}
\end{equation}
The electric potential and tangential electric field are continuous at $r=a$,
\begin{equation}
 \jump{\phi}=0,
 \qquad
 \jump{\bE\cdot\bet}=0,
 \label{eq:dim_electric_continuity}
\end{equation}
and the free surface-charge density $q(\theta,t)$ is determined by Gauss's law,
\begin{equation}
 q=\jump{\epsilon\bE}\cdot\bn
   =\epsilon^+E_n^+-\epsilon^-E_n^-, \qquad r=a.
 \label{eq:dim_gauss}
\end{equation}
Here $E_n^\pm=\bE^\pm\cdot\bn$. The charge evolves through Ohmic currents from the two bulk phases, convection by the interfacial flow and a surface-diffusion term,
\begin{equation}
 \frac{\partial q}{\partial t}
 +\jump{\sigma\bE}\cdot\bn
 +\grad_s\cdot(q\bu)
 =D_q^{\mathrm{dim}}\nabla_s^2q,
 \qquad r=a.
 \label{eq:dim_charge_general}
\end{equation}
Here $\grad_s=(\bI-\bn\bn)\cdot\grad$ and $D_q^{\mathrm{dim}}$ is the dimensional surface-charge diffusivity. Surface-charge diffusion has been retained in earlier electrohydrodynamic calculations, including those of \citet{Collins2013}. In the circular-drop problem, weak charge diffusion also regularises the sharp charge-density structures associated with the singular diffusionless limit.\citep{PengSchnitzer2025} We retain this term to resolve the charge distribution at finite diffusivity under strong convection. On the fixed circle, Eq.~\eqref{eq:dim_charge_general} becomes
\begin{equation}
 \frac{\partial q}{\partial t}
 +\sigma^+E_n^+-\sigma^-E_n^-
 +\frac{1}{a}\frac{\partial}{\partial\theta}(q u_\theta)
 =\frac{D_q^{\mathrm{dim}}}{a^2}\frac{\partial^2q}{\partial\theta^2}.
 \label{eq:dim_charge_circle}
\end{equation}

\subsection{Creeping flow and interfacial stresses}

Fluid inertia is neglected because the Reynolds numbers of the small drops in the viscous liquids of interest are typically very small. The velocity and pressure fields therefore obey the Stokes equations
\begin{equation}
 -\grad p^\pm+\mu^\pm\nabla^2\bu^\pm=\bm{0},
 \qquad
 \grad\cdot\bu^\pm=0
 \quad \text{in }V^\pm,
 \label{eq:dim_stokes}
\end{equation}
with hydrodynamic stress
\begin{equation}
 \bm{T}_H^\pm=-p^\pm\bI
 +\mu^\pm\left[\grad\bu^\pm+(\grad\bu^\pm)^T\right].
 \label{eq:dim_hydrostress}
\end{equation}
The exterior fluid is quiescent at infinity and the interior velocity is regular at the origin. Velocity is continuous across the interface, and the prescribed geometry imposes no penetration:
\begin{equation}
 u_r^+=u_r^-=0,
 \qquad
 u_\theta^+=u_\theta^-=u_\theta(\theta,t), \qquad r=a.
 \label{eq:dim_velocity_bc}
\end{equation}
The electric stress is described by the Maxwell tensor
\begin{equation}
 \bm{T}_E^\pm=\epsilon^\pm
 \left(\bE^\pm\bE^\pm-\frac{1}{2}|\bE^\pm|^2\bI\right).
 \label{eq:dim_maxwell}
\end{equation}
For a freely deforming interface, normal traction balance determines the evolving shape. Here the shape is prescribed and the normal velocity is set to zero instead. We represent the normal traction required to maintain this constraint by $\Lambda(\theta,t)$, so that the total traction balance is
\begin{equation}
 \jump{\bm{T}_H+\bm{T}_E}\cdot\bn
 =\gamma\kappa\bn-\grad_s\gamma+\Lambda\bn,
 \qquad r=a,
 \label{eq:dim_full_stress}
\end{equation}
where $\kappa=1/a$. Projecting the same traction balance onto the tangent $\bet$, and using continuity of the tangential electric field together with Eq.~\eqref{eq:dim_gauss}, gives
\begin{equation}
 \bet\cdot\jump{\bm{T}_H}\cdot\ber
 +qE_\theta
 +\frac{1}{a}\frac{\partial\gamma}{\partial\theta}=0,
 \qquad r=a.
 \label{eq:dim_tangential_stress}
\end{equation}
Equation~\eqref{eq:dim_tangential_stress} displays the three tangential stresses central to the present problem: the hydrodynamic stress $\bet\cdot\jump{\bm{T}_H}\cdot\ber$, the electric shear stress $qE_\theta$, and the Marangoni stress $a^{-1}\partial\gamma/\partial\theta$.

\subsection{Surfactant transport and equation of state}

The interface is coated by surfactant of concentration $\Gamma(\theta,t)$. We consider both an insoluble layer and a soluble-surfactant exchange model in which the interface exchanges surfactant with a prescribed, spatially uniform bulk reservoir of concentration $c_b$. For finite-site adsorption with rate constant $k_a$ and desorption rate $k_d$, the net flux into the interface is
\begin{equation}
 J=k_a c_b(\Gamma_\infty-\Gamma)-k_d\Gamma
     =k_{\mathrm{ex}}(\Gamma_{\mathrm{eq}}-\Gamma),
 \label{eq:dim_exchange}
\end{equation}
where
\begin{equation}
 k_{\mathrm{ex}}=k_a c_b+k_d,
 \qquad
 \Gamma_{\mathrm{eq}}
 =\Gamma_\infty\frac{k_a c_b}{k_a c_b+k_d}.
 \label{eq:dim_exchange_defs}
\end{equation}
Thus the soluble model represents kinetic exchange with a well-mixed reservoir; no bulk surfactant concentration field is needed in this model. The insoluble limit is recovered by setting $k_{\mathrm{ex}}=0$ while retaining the chosen mean surface concentration.

Surface advection, surface diffusion and exchange with the well-mixed bulk reservoir give
\begin{equation}
 \frac{\partial\Gamma}{\partial t}
 +\grad_s\cdot(\Gamma\bu)
 =D_\Gamma^{\mathrm{dim}}\nabla_s^2\Gamma+J,
 \qquad r=a,
 \label{eq:dim_surfactant_general}
\end{equation}
where $D_\Gamma^{\mathrm{dim}}$ is the dimensional surface diffusivity of the surfactant. On the circle, this reduces to
\begin{equation}
 \frac{\partial\Gamma}{\partial t}
 +\frac{1}{a}\frac{\partial}{\partial\theta}(\Gamma u_\theta)
 =\frac{D_\Gamma^{\mathrm{dim}}}{a^2}\frac{\partial^2\Gamma}{\partial\theta^2}
 +k_{\mathrm{ex}}(\Gamma_{\mathrm{eq}}-\Gamma).
 \label{eq:dim_surfactant_circle}
\end{equation}
Its angular mean satisfies
\begin{equation}
 \frac{d\langle\Gamma\rangle}{dt}
 =k_{\mathrm{ex}}(\Gamma_{\mathrm{eq}}-\langle\Gamma\rangle).
 \label{eq:dim_surf_mean}
\end{equation}
We initialise the mean concentration at $\Gamma_{\mathrm{eq}}$, so the mean remains fixed even though adsorption and desorption may occur locally.
For a finite-coverage monolayer we use a Langmuir equation of state,
\begin{equation}
 \gamma=\gamma_0+R_gT\Gamma_\infty
 \ln\left(1-\frac{\Gamma}{\Gamma_\infty}\right),
 \label{eq:dim_langmuir}
\end{equation}
where $R_g$ is the gas constant, $T$ the absolute temperature, $\Gamma_\infty$ the maximum surface concentration and $\gamma_0$ the clean-interface surface tension. We denote by $\gamma_{\mathrm{eq}}$ the surface tension evaluated at $\Gamma=\Gamma_{\mathrm{eq}}$.

\section{Nondimensional formulation}
\label{sec:nondimensional}

\subsection{Scales and dimensionless groups}

Lengths are scaled with $a$ and time with the exterior charge-relaxation time
\begin{equation}
 \tau^+=\frac{\epsilon^+}{\sigma^+}.
\end{equation}
The electroviscous velocity scale is
\begin{equation}
 U_E=\frac{\epsilon^+E_0^2a}{\mu^+}.
\end{equation}
This scale follows from balancing electric and viscous stresses. Although it depends on the applied field, we retain it for consistency with previous electrohydrodynamic studies. For comparisons between different applied fields, we also introduce the fixed electroviscous velocity
\begin{equation}
 U_{c,s}=\frac{\epsilon^+E_{c,s}^{\,2}a}{\mu^+},
 \qquad
 \widetilde{\bu}=\left(\frac{E_0}{E_{c,s}}\right)^2\bu.
 \label{eq:fixed_velocity_scale}
\end{equation}
Here $E_0$ is the applied field and $E_{c,s}$ is the critical field for Quincke rotation of the reference rigid cylinder, specified in Eq.~\eqref{eq:rigid_critical_field}. The tilded velocity is nondimensionalised by $U_{c,s}$, whereas $\bu$ is nondimensionalised by $U_E$. The fixed scale is used in Fig.~\ref{fig:low_diffusion_flow}. Figures~S1--S7 of the Supplementary Material reproduce Figs.~2, 3, 5, 8, 11, 12 and 13, respectively, with velocities, rotation rates and strain intensities expressed using the corresponding fixed reference scales. The electrohydrodynamic flow time associated with $U_E$ is
\begin{equation}
 \tau_{\rm EHD}=\frac{a}{U_E}
 =\frac{\mu^+}{\epsilon^+E_0^2}.
 \label{eq:tau_EHD}
\end{equation}
Electric potential, electric field and surface charge are scaled with $E_0a$, $E_0$ and $\epsilon^+E_0$, respectively. Pressure and both hydrodynamic and electric stresses are scaled with $\epsilon^+E_0^2$. Surface tension is scaled with $\gamma_{\mathrm{eq}}$ and surfactant concentration with $\Gamma_{\mathrm{eq}}$. The same symbols are used below for the corresponding dimensionless variables, and the interface is at $r=1$. The material ratios are
\begin{equation}
 R=\frac{\sigma^+}{\sigma^-},
 \qquad
 S=\frac{\epsilon^-}{\epsilon^+},
 \qquad
 M=\frac{\mu^-}{\mu^+}.
 \label{eq:ratios}
\end{equation}
The inner charge-relaxation time is $\tau^-=\epsilon^-/\sigma^-$. Hence
\begin{equation}
 RS=\frac{\tau^-}{\tau^+}
 =\frac{\epsilon^-/\sigma^-}{\epsilon^+/\sigma^+},
 \label{eq:RS_tau}
\end{equation}
so $RS$ directly compares charge relaxation inside and outside the drop. The Quincke rotational instability requires $RS>1$, or equivalently $\tau^->\tau^+$. The electric Reynolds number compares the charge-relaxation time of the outer liquid with the electroviscous advection time,
\begin{equation}
 \ReE=\frac{\tau^+U_E}{a}
 =\frac{(\epsilon^+)^2E_0^2}{\sigma^+\mu^+}.
 \label{eq:ReE}
\end{equation}
The dimensionless surface diffusivities and exchange rate are
\begin{equation}
 \Dq=\frac{D_q^{\mathrm{dim}}\tau^+}{a^2},
 \qquad
 \DG=\frac{D_\Gamma^{\mathrm{dim}}\tau^+}{a^2},
 \qquad
 \Ks=k_{\mathrm{ex}}\tau^+.
 \label{eq:diffusion_groups}
\end{equation}
We also define the electric capillary number and surfactant elasticity number,
\begin{equation}
 \CaE=\frac{\epsilon^+E_0^2a}{\gamma_{\mathrm{eq}}},
 \qquad
 \El=\frac{R_gT\Gamma_{\mathrm{eq}}}{\gamma_{\mathrm{eq}}},
 \label{eq:CaEl}
\end{equation}
and their ratio, the Marangoni number,
\begin{equation}
 \Ma=\frac{\El}{\CaE}
 =\frac{R_gT\Gamma_{\mathrm{eq}}}{\epsilon^+E_0^2a}.
 \label{eq:Ma}
\end{equation}
Here $\Ma$ denotes the Marangoni number, and must not be confused with the Mason number used for the viscous-to-electric stress ratio in earlier electrohydrodynamic studies.\citep{DasSaintillan2013,DasSaintillan2017small,DasSaintillan2017,DasSaintillan2021} The fixed-circle approximation is motivated by $\CaE\ll1$, provided that the local surface tension remains positive and does not become asymptotically small relative to $\gamma_{\mathrm{eq}}$.  Circularity is imposed exactly in the present model, and the deformation correction is not computed. A useful physical interpretation follows by defining the Marangoni--viscous time
\begin{equation}
 \tau_{\rm Ma}=\frac{\mu^+a}{R_gT\Gamma_{\mathrm{eq}}},
 \label{eq:tau_Ma}
\end{equation}
so that $\Ma=\tau_{\rm EHD}/\tau_{\rm Ma}$. At fixed material properties, drop size and reference surface concentration, $\ReE\propto E_0^2$ whereas $\Ma\propto E_0^{-2}$. It is therefore convenient to introduce the field-independent surfactant-strength parameter
\begin{equation}
 B\equiv \ReE\Ma
 =\frac{\epsilon^+R_gT\Gamma_{\mathrm{eq}}}{\sigma^+\mu^+a}
 =\frac{\tau^+}{\tau_{\rm Ma}}.
 \label{eq:B}
\end{equation}
Thus a sweep in applied field for a single drop--fluid system is made at fixed $B$, with $\Ma=B/\ReE$; keeping $\Ma$ fixed while
changing $\ReE$ would instead change the material system. The rigid-cylinder critical field, including surface-charge diffusion, is 
\begin{equation}
 E_{c,s}
 =
 \left[
 \frac{
 \mu^+\left[
 \sigma^++\sigma^-+
 (\epsilon^++\epsilon^-)\,D_q^{\mathrm{dim}}/a^2
 \right]^2
 }{
 \epsilon^+(\epsilon^-\sigma^+-\epsilon^+\sigma^-)
 }
 \right]^{1/2},
 \label{eq:rigid_critical_field}
\end{equation}
where $D_q^{\mathrm{dim}}$ denotes the dimensional surface-charge diffusivity. This threshold exists when $\epsilon^-\sigma^+>\epsilon^+\sigma^-$, equivalently $RS>1$. The dimensionless applied field is then
\begin{equation}
 \frac{E_0}{E_{c,s}}
 =
 \left(
 \frac{\ReE}{\mathrm{Re}_{E,c,s}}
 \right)^{1/2},
 \label{eq:Ec_scale}
\end{equation}
where $\mathrm{Re}_{E,c,s}$ is the corresponding rigid-cylinder critical electric Reynolds number. For the baseline electrical system, $R=S=2$ and $\Dq=10^{-3}$, $\mathrm{Re}_{E,c,s}=1.5060$. The same rigid-cylinder critical field $E_{c,s}$ is used as the reference throughout.

We denote by $E_{c,d}$ the field at which the non-rotating Taylor state first becomes unstable to Quincke rotation. For a subcritical transition, stable rotating states also exist below $E_{c,d}$, down to the saddle-node shown in Fig.~\ref{fig:rotation_branches}. The corresponding critical Reynolds numbers are $\mathrm{Re}_{E,c,s}$ and $\mathrm{Re}_{E,c,d}$. The threshold ratio \(E_{c,d}/E_{c,s}\) depends on the electrical material ratios, the viscosity ratio, the surfactant coverage and elasticity, and the charge and surfactant transport parameters.

Finally, the equilibrium surface coverage is
\begin{equation}
 \chi=\frac{\Gamma_{\mathrm{eq}}}{\Gamma_\infty},
 \qquad 0\leq\chi<1.
 \label{eq:chi}
\end{equation}
For an insoluble monolayer, $\Gamma_{\mathrm{eq}}$ is the prescribed mean surface concentration. Increasing the mean concentration of a given surfactant at fixed $\Gamma_\infty$ changes both $B$ and $\chi$; the fixed-$\chi$ sweeps therefore do not represent variation of surfactant concentration alone. The model parameters $B$ and $\chi$ can be specified independently, and choosing $B=1$ does not determine the coverage.
The quantity $B$ is an electric field-independent measure of surfactant strength. The dimensional Gibbs (or Marangoni) elasticity, evaluated at the mean concentration, is
\begin{equation}
 E_s
 =-\Gamma_{\mathrm{eq}}
 \left.
 \frac{d\gamma^{\mathrm{dim}}}{d\Gamma^{\mathrm{dim}}}
 \right|_{\Gamma^{\mathrm{dim}}=\Gamma_{\mathrm{eq}}}
 =\frac{R_gT\Gamma_{\mathrm{eq}}}{1-\chi},
 \qquad
 \frac{E_s}{\mu^+a/\tau^+}=\frac{B}{1-\chi}.
 \label{eq:Gibbs_elasticity}
\end{equation}
Thus $B/(1-\chi)$, rather than $B$ alone, measures the surface elasticity at the mean concentration in units set by viscous stresses and the charge-relaxation time. \citep{ManikantanSquires2020,Champougny2015} The dimensionless Langmuir equation of state is
\begin{equation}
 \gamma
 =1+\frac{\El}{\chi}
 \ln\left(\frac{1-\chi\Gamma}{1-\chi}\right),
 \qquad
 \frac{\partial\gamma}{\partial\theta}
 =-\frac{\El}{1-\chi\Gamma}\frac{\partial\Gamma}{\partial\theta}.
 \label{eq:nd_langmuir}
\end{equation}
The equation of state requires $0\leq\chi\Gamma<1$. For dilute surface coverage, with $\chi\ll1$ and $\Gamma=O(1)$, this reduces to the linear equation of state $\gamma\simeq1-\El(\Gamma-1)$. Table~S1 in the Supplementary Material gives an illustrative dimensional realisation of selected parameters for stated fluid properties, temperature and drop radius; the accompanying text gives the conversion relations for the transport parameters.
\begin{figure*}[tp]
 \centering
 \includegraphics[width=0.7\textwidth]{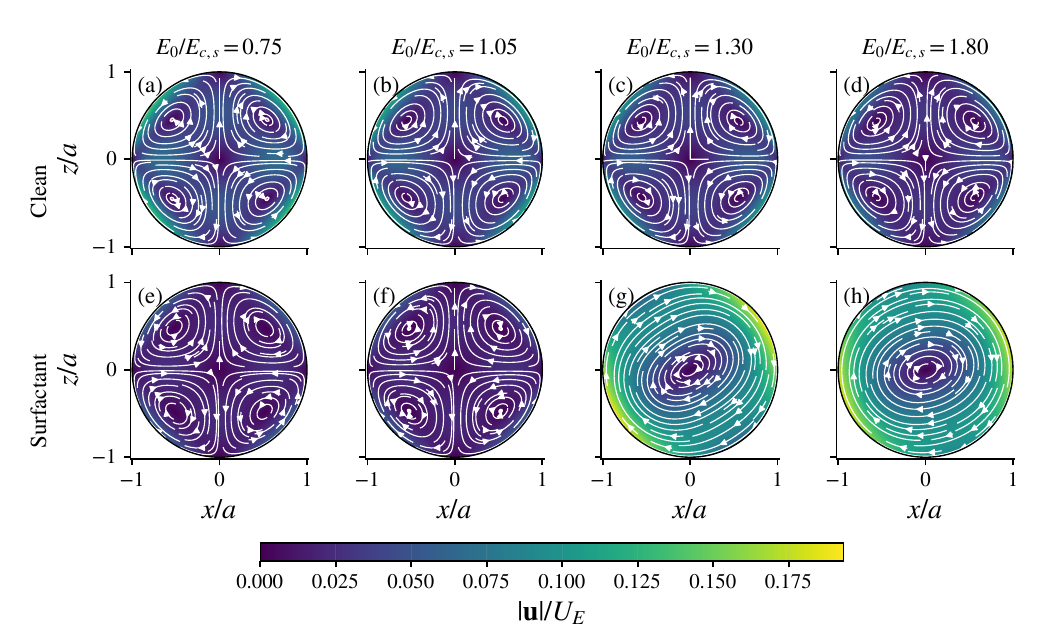}
 \caption{Interior flow for clean (top) and surfactant-covered (bottom) drops at matched applied fields. The surfactant case has $R=S=2$, $M=1$, $B=1$, $\DG=0.1$, $\Dq=10^{-3}$ and $\Ks=0$. Colour shows the dimensionless speed. The clean Taylor profiles at $E_0/E_{c,s}=1.30$ and $1.80$ use $N=4096$. The clean drop remains in the Taylor regime over the displayed range, whereas surfactant strongly suppresses the Taylor circulation before the onset of Quincke rotation and the covered drop subsequently enters the Quincke regime.}
 \label{fig:flow_fields}
\end{figure*}

\subsection{Dimensionless equations and boundary conditions}

The electric potentials satisfy
\begin{equation}
 \nabla^2\phi^-=0, \qquad r<1,
 \label{eq:nd_laplace_in}
\end{equation}
\begin{equation}
 \nabla^2\phi^+=0, \qquad r>1,
 \label{eq:nd_laplace_out}
\end{equation}
with
\begin{equation}
 \phi^+ \longrightarrow - r\cos\theta
 \qquad r\longrightarrow\infty,
 \label{eq:nd_farfield}
\end{equation}
and, at $r=1$,
\begin{equation}
 \phi^+=\phi^-,
 \qquad
 q=E_r^+-SE_r^-.
 \label{eq:nd_electric_bc}
\end{equation}
The dimensionless surface-charge balance is
\begin{equation}
 \frac{\partial q}{\partial t}
 +\ReE\frac{\partial}{\partial\theta}(q u_\theta)
 +E_r^+-\frac{1}{R}E_r^-
 =\Dq\frac{\partial^2q}{\partial\theta^2},
 \qquad r=1.
 \label{eq:nd_charge}
\end{equation}
For an initially neutral drop, integration of Eq.~\eqref{eq:nd_charge} together with Gauss's law preserves zero net charge,
\begin{equation}
 \int_0^{2\pi}q(\theta,t)\,d\theta=0.
 \label{eq:nd_neutrality}
\end{equation}

The Stokes equations become
\begin{align}
 -\grad p^+ +\nabla^2\bu^+&=\bm{0},
 &\grad\cdot\bu^+&=0,
 &&r>1,
 \label{eq:nd_stokes_plus}\\
 -\grad p^- +M\nabla^2\bu^-&=\bm{0},
 &\grad\cdot\bu^-&=0,
 &&r<1.
 \label{eq:nd_stokes_minus}
\end{align}
The exterior flow tends to zero at infinity, and the interfacial conditions are
\begin{equation}
 u_r^+=u_r^-=0,
 \qquad
 u_\theta^+=u_\theta^-=u_\theta(\theta,t),
 \qquad r=1.
 \label{eq:nd_velocity_bc}
\end{equation}
The dimensionless hydrodynamic and Maxwell stresses are
\begin{align}
 \bm{T}_H^+&=-p^+\bI+\grad\bu^++(\grad\bu^+)^T,
 \label{eq:nd_TH_plus}\\
 \bm{T}_H^-&=-p^-\bI+M\left[\grad\bu^-+(\grad\bu^-)^T\right],
 \label{eq:nd_TH_minus}\\
 \bm{T}_E^+&=\bE^+\bE^+-\frac12|\bE^+|^2\bI,
 \label{eq:nd_TE_plus}\\
 \bm{T}_E^-&=S\left(\bE^-\bE^--\frac12|\bE^-|^2\bI\right).
 \label{eq:nd_TE_minus}
\end{align}
The complete interfacial traction balance is
\begin{equation}
 \jump{\bm{T}_H+\bm{T}_E}\cdot\ber
 =\frac{1}{\CaE}\left(\gamma\ber-\grad_s\gamma\right)
 +\Lambda\ber,
 \qquad r=1.
 \label{eq:nd_full_stress}
\end{equation}
Projecting the dimensionless traction balance onto the tangent and using the Langmuir equation of state gives
\begin{equation}
 \jump{T_{H,r\theta}}+qE_\theta
 -\frac{\Ma}{1-\chi\Gamma}\frac{\partial\Gamma}{\partial\theta}=0,
 \qquad r=1.
 \label{eq:nd_tangential_langmuir}
\end{equation}
Because the interface radius is prescribed, the normal component of Eq.~\eqref{eq:nd_full_stress} determines only the constraint traction $\Lambda$. The equations for tangential surface motion contain $\El$ and $\CaE$ only through their ratio $\Ma$.

The surfactant transport equation is
\begin{equation}
 \frac{\partial\Gamma}{\partial t}
 +\ReE\frac{\partial}{\partial\theta}(\Gamma u_\theta)
 =\DG\frac{\partial^2\Gamma}{\partial\theta^2}+\Ks(1-\Gamma),
 \qquad r=1,
 \label{eq:nd_surfactant}
\end{equation}
with
\begin{equation}
 \frac{1}{2\pi}\int_0^{2\pi}\Gamma(\theta,t)\,d\theta=1
 \label{eq:nd_surf_mean}
\end{equation}
for the equilibrium-mean states considered here. Setting $\Ks=0$ recovers an insoluble surfactant with conserved mean concentration; setting $\Ma=0$ removes the Marangoni feedback and recovers the clean-interface electrohydrodynamic problem. Equations~\eqref{eq:nd_laplace_in}--\eqref{eq:nd_surf_mean} define the nonlinear circular-drop model studied below. Because the interface is circular, the bulk electric and Stokes problems can be eliminated analytically, leaving a closed interfacial problem for $q(\theta,t)$ and $\Gamma(\theta,t)$. The numerical methods and validation checks are described in the Supplementary Material.

\section{Results and discussion}
\label{sec:results}

We first examine the steady branches and the recovery of circulation at rotational onset, then quantify the flow and its temporal evolution, and finally examine the parameter dependence of the onset and compare surface diffusion with bulk exchange.

\subsection{Parameter families and representative interfacial states}
\label{sec:results_parameters}

We first consider an insoluble surfactant ($\Ks=0$) and then return to soluble exchange in Sec.~\ref{sec:solubility_results}. Unless otherwise stated, the electrical ratios are $R=S=2$, the surface-charge diffusivity is $\Dq=10^{-3}$ and the equilibrium coverage is $\chi=0.2$. The baseline surfactant-covered family has $M=1$ and $B=1$, while the baseline surface diffusivity is varied over $\DG=1,\,0.1,\,0.01$ and $0.001$. The rotation-branch comparison additionally includes $\DG=10$. Clean-interface calculations use the same electrical and viscous material properties with the Marangoni coupling removed ($\Ma=0$). Additional sweeps vary $M$, $B$ and $\DG$ independently in order to separate viscous, elastic and surfactant-transport effects. The reported thresholds and branches refer to the finite charge diffusivity $\Dq=0.001$.

It is important to distinguish changing the electric field in one physical drop--fluid system from varying dimensionless parameters independently. From Eqs.~\eqref{eq:ReE}, \eqref{eq:Ma} and \eqref{eq:B}, varying only $E_0$ for a given drop--fluid pair increases $\ReE$ while decreasing $\Ma$ so that $B=\ReE\Ma$ remains constant. All field-dependent branches shown against $E_0/E_{c,s}$ therefore follow this fixed-material path. Surface diffusion $\DG$ and exchange $\Ks$ are also defined using the outer charge-relaxation time and remain fixed when only the field is varied. The same rigid-cylinder value $E_{c,s}$ is used as the reference field throughout.

The examples below are supplemented by sweeps of viscosity ratio, surfactant strength and surface diffusivity in Figs.~\ref{fig:onset_strength}--\ref{fig:diffusion_cuts}. These calculations establish the parameter dependence within the chosen electrical system, $R=S=2$, at $\chi=0.2$ and $\Dq=10^{-3}$. The comparison of surface diffusion with exchange is examined separately in Sec.~\ref{sec:solubility_results}.

\subsection{Immobilisation of Taylor circulation and remobilisation on rotation}
\label{sec:immobilisation_results}

The effect of surfactant is already clear in the interior flow fields of Fig.~\ref{fig:flow_fields}. The top row shows the clean drop at four applied electric fields. Throughout the displayed range the clean $M=1$ drop remains in the non-rotating Taylor regime and retains the familiar Taylor circulation. At the same electric fields, an insoluble surfactant with $B=1$ and $\DG=0.1$ strongly suppresses the Taylor circulation before the onset of Quincke rotation. The electrically driven surface motion advects surfactant, and the resulting Marangoni stress opposes the electric shear stress. Once the surfactant-covered drop begins to rotate, a strong mean azimuthal flow appears and the straining circulation also increases. Quincke rotation therefore remobilises the nearly immobilised interface and restores a substantial straining circulation. We quantify the net rotation by the mean angular velocity
\begin{equation}
 \Omega=\langle u_\theta\rangle.
 \label{eq:Omega_results}
\end{equation}
Because the drop is fixed and centred, translation is excluded; the zeroth Fourier mode of $u_\theta$ therefore represents rigid rotation.

\begin{figure}[tp]
 \centering
 \includegraphics[width=0.95\columnwidth]{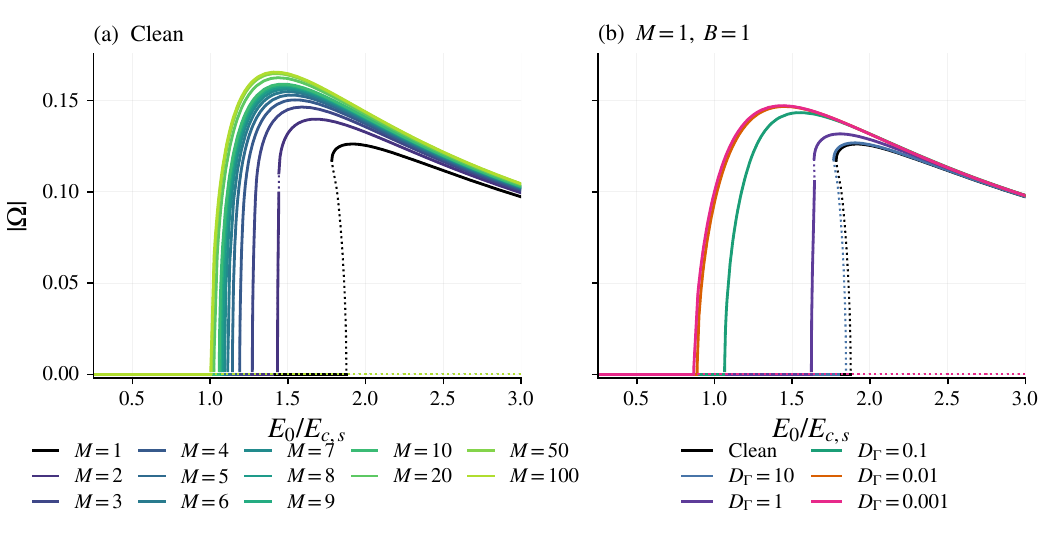}
 \caption{Steady rotation branches. Left: clean drops for viscosity ratios $M=1,\ldots,10,20,50,100$. Right: $M=1$, $B=1$ with insoluble surfactant at the indicated diffusivities, together with the clean reference. The field is normalised by the common rigid-cylinder value $E_{c,s}$. Solid lines denote states classified as stable to perturbations preserving the imposed half-turn symmetries; dotted lines denote unstable or marginal continuations.}
 \label{fig:rotation_branches}
\end{figure}

The steady branches in Fig.~\ref{fig:rotation_branches} show that surfactant changes the critical field for Quincke rotation, the supercritical or subcritical character of the rotational bifurcation, and the shape of the rotating branch. Here, bistability denotes the coexistence, at the same applied field and parameter values, of two stable states with different rotation magnitudes: either a non-rotating Taylor state and a rotating state, or two rotating states with different amplitudes. For a clean drop, reducing $M$ raises the critical field above the rigid-cylinder value and, at sufficiently low viscosity ratio, produces a subcritical rotating branch whose unstable segment extends towards lower electric fields and joins a stable rotating segment at a saddle-node bifurcation, creating the bistable interval found in the clean circular-drop problem.\citep{PengSchnitzer2025} This coexistence is qualitatively consistent with the experimental hysteresis reported by \citet{SalipanteVlahovska2010}. The right panel fixes $M=1$ and $B=1$ and varies $\DG$. At large $\DG$, the rotating branch remains close to the clean-drop branch. Decreasing $\DG$ progressively lowers the critical field for Quincke rotation. At $\DG=0.01$ and $0.001$ rotation begins below the rigid-cylinder reference field $E_{c,s}$. In addition to suppressing Taylor circulation, surfactant transport changes the rotational instability: for this family, reducing $\DG$ lowers the critical field and changes the onset from subcritical in the clean and sufficiently high-diffusion cases ($\DG=10$) to supercritical at lower diffusion. At $\DG=1$, the onset is supercritical, but the rotating branch develops an S-shaped segment at finite rotation amplitude, bounded by two saddle-node (fold) bifurcations. This produces a narrow field interval containing two stable rotating states of different amplitudes, separated by an unstable branch. A magnified view is shown in Fig.~S8 of the Supplementary Material.

\begin{figure}[tp]
 \centering
 \includegraphics[width=0.95\columnwidth]{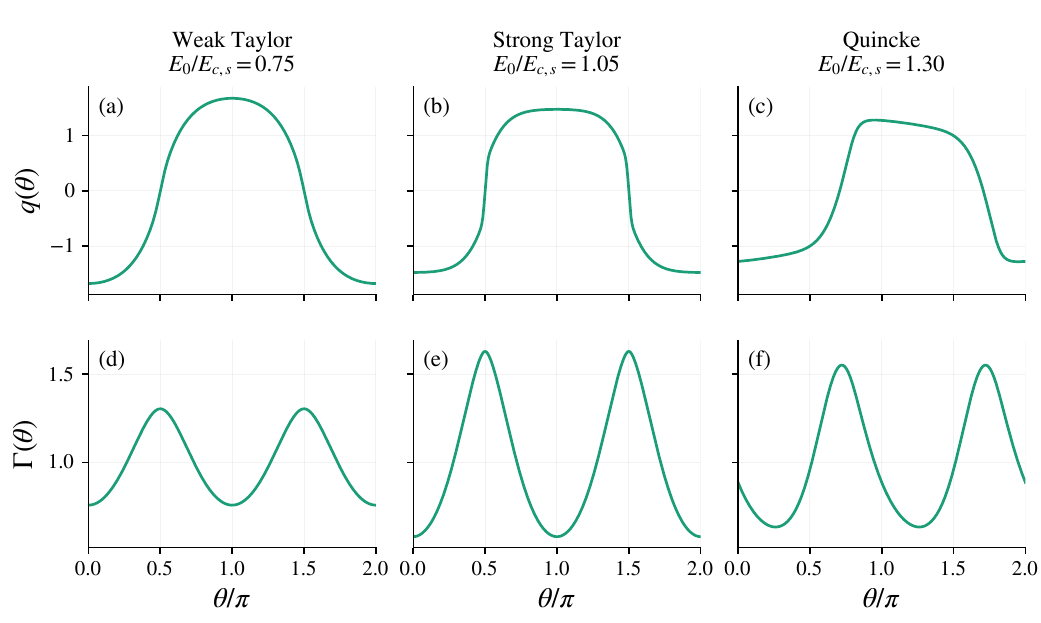}
 \caption{Surface charge $q(\theta)$ (top row) and surfactant concentration $\Gamma(\theta)$ (bottom row) for the baseline surfactant-covered drop with $R=S=2$, $M=1$, $B=1$, $\DG=0.1$, $\Dq=10^{-3}$ and $\Ks=0$. From left to right: weak Taylor regime, $E_0/E_{c,s}=0.75$; strong pre-onset Taylor regime, $E_0/E_{c,s}=1.05$; and Quincke regime, $E_0/E_{c,s}=1.30$.}
 \label{fig:surface_profiles}
\end{figure}

Figure~\ref{fig:surface_profiles} shows the interfacial fields underlying these changes for the baseline surfactant-covered drop. In the weak Taylor regime the charge and surfactant distributions are smooth and retain the reflection symmetries of the applied field. Closer to onset, charge convection produces a much sharper charge profile and the surfactant redistribution becomes stronger. In the Quincke regime the charge distribution is tilted relative to the applied field and the surfactant pattern is shifted around the interface, consistent with the broken reflection symmetry and finite mean rotation. These profiles provide a direct view of the two advected interfacial fields before introducing global measures of their effect on the flow.

Movies~S1 and S2 show the development of Taylor circulation and Quincke rotation for a clean drop. Movies~S3 and S4 show the corresponding evolution with a weakly diffusing insoluble surfactant, including immobilisation and subsequent remobilisation in Movie~S4. The movie parameters and initial conditions are given in the Supplementary Material.

\subsection{Measures of interfacial and interior flow}
\label{sec:mobility_diagnostics}

We quantify interfacial motion using the RMS tangential speed and the RMS speed remaining after subtraction of mean rotation,
\begin{equation}
 U_{\rm rms}=\langle u_\theta^2\rangle^{1/2},
 \qquad
 U_{\rm str}=\langle(u_\theta-\Omega)^2\rangle^{1/2}.
 \label{eq:surface_mobility_results}
\end{equation}
A signed average cannot measure the Taylor circulation because oppositely directed surface motions cancel. We decompose $u_\theta=\Omega+u_{\rm str}$ with $\langle u_{\rm str}\rangle=0$. The mean tangential surface velocity generates rigid-body rotation inside the drop, $\bu_{\rm rot}=\Omega r\bet$. We call the remaining non-rigid flow the straining circulation; it can contain local vorticity as well as strain. The interfacial measure $U_{\rm str}$ quantifies its surface-speed amplitude and includes all resolved nonzero velocity harmonics. The decomposition gives $U_{\rm rms}^2=\Omega^2+U_{\rm str}^2$.

\begin{figure}[tp]
 \centering
 \includegraphics[width=0.95\columnwidth]{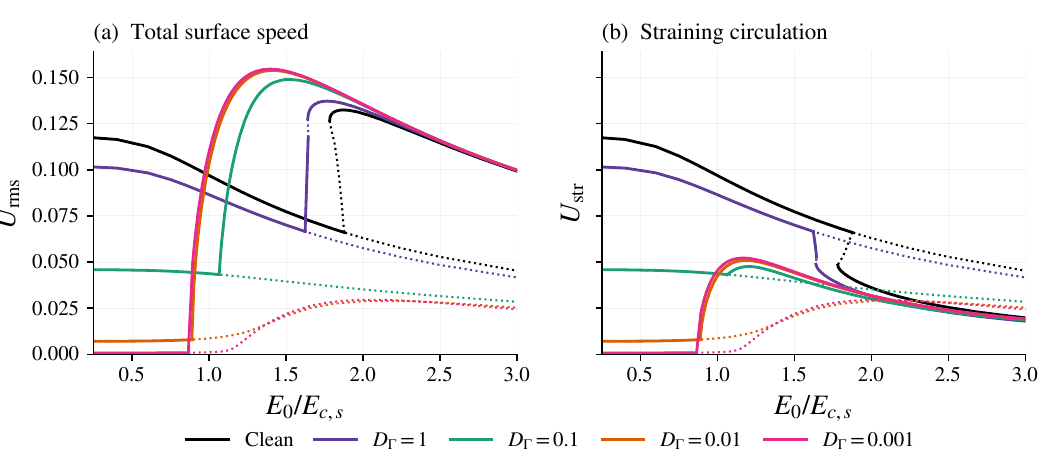}
 \caption{Measures of interfacial flow for $R=S=2$, $M=1$ and $B=1$. Left: total RMS tangential speed $U_{\rm rms}$. Right: straining RMS speed $U_{\rm str}$ after subtracting the mean rotation. Solid and dotted portions follow the same stability convention as Fig.~\ref{fig:rotation_branches}.}
 \label{fig:mobility}
\end{figure}

Figure~\ref{fig:mobility} separates the total surface speed from its straining component. In the Taylor regime $\Omega=0$, so the two measures coincide. Low diffusion can suppress this interfacial flow by more than an order of magnitude relative to the clean interface. At the onset of Quincke rotation $U_{\rm rms}$ rises sharply because of the newly acquired mean rotation, but the change is not purely rigid-body motion: $U_{\rm str}$ also recovers. This is clearest for $\DG=0.01$ and $0.001$: the Taylor circulation is nearly arrested before onset, but a finite straining component reappears after rotation begins. The rotational instability therefore restores both mean azimuthal motion and a substantial circulation suppressed by Marangoni stresses.

These speeds are nondimensionalised by the field-dependent electroviscous velocity $U_E=\epsilon^+E_0^2a/\mu^+$. A decrease in $U_{\rm rms}$ or $U_{\rm str}$ therefore means a decrease relative to this scale and need not imply slower dimensional speeds. For example, on the clean $M=1$ Taylor branch, $U_{\rm rms}$ decreases from $0.1164$ at $E_0/E_{c,s}=0.4$ to $0.0969$ at $1.0$, whereas the same speeds on the fixed scale $U_{c,s}$ increase from $0.0186$ to $0.0969$. Charge convection redistributes the electric traction and reduces the large-scale circulation relative to $U_E$. Its spatial distribution matters as well as its magnitude: finer angular variations produce weaker velocity responses. Indeed, the RMS electric traction in units of $\epsilon^+E_0^2$ increases slightly from $0.941$ to $0.967$ between these states, while the quadrupolar contribution to the surface speed decreases from $0.1160$ to $0.0938$. For a surfactant-covered interface, Marangoni traction also contributes to the stress balance. Corresponding speed plots using the fixed scale $U_{c,s}$ are provided in the Supplementary Material.

\begin{figure}[tp]
 \centering
 \includegraphics[width=0.95\columnwidth]{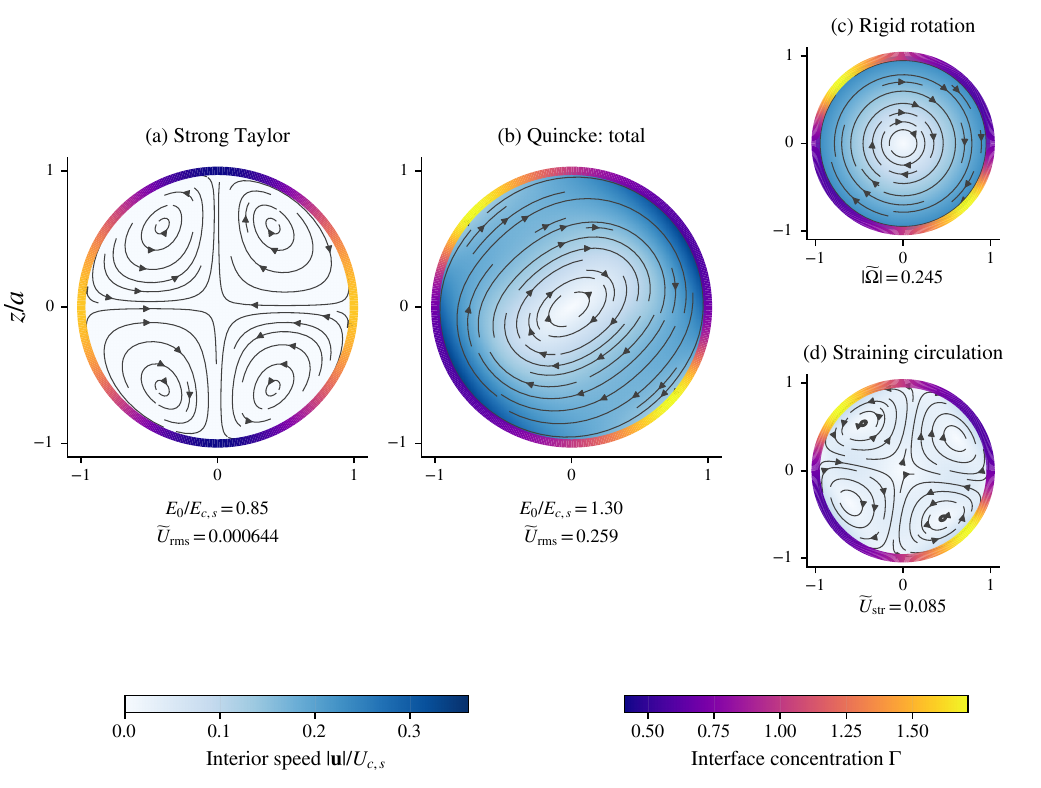}
 \caption{Low-diffusion immobilisation and recovery for $R=S=2$, $M=B=1$, $\chi=0.2$, $\DG=10^{-3}$, $\Dq=10^{-3}$ and $\Ks=0$. (a) Strong Taylor state at $E_0/E_{c,s}=0.85$. (b) Total Quincke flow at $1.30$, decomposed into (c) rigid rotation and (d) straining circulation. Interior shading shows speed on one common fixed-reference scale $U_{c,s}$; streamlines show direction, not magnitude. The interface colour shows the dimensionless concentration $\Gamma$ on a second common scale; surfactant is not a bulk field. The same Quincke concentration is shown in (b)--(d). Annotations give the total RMS speed $\widetilde U_{\rm rms}$ in (a,b), the rotational surface speed $|\widetilde\Omega|$ in (c), and the straining RMS speed $\widetilde U_{\rm str}$ in (d), where each tilded quantity is multiplied by $(E_0/E_{c,s})^2$ relative to its $U_E$ normalisation. In (a), $\widetilde U_{\rm rms}=\widetilde U_{\rm str}$ because $\Omega=0$.}
 \label{fig:low_diffusion_flow}
\end{figure}

Figure~\ref{fig:low_diffusion_flow} compares the interfacial concentration and interior flow before and after the onset of rotation at $\DG=10^{-3}$. At $E_0/E_{c,s}=0.85$, just below $E_{c,d}/E_{c,s}=0.8663$, surfactant accumulates towards the equator and the Taylor circulation is nearly arrested. At $E_0/E_{c,s}=1.30$, the surfactant pattern shifts around the interface, while both mean rotation and straining circulation are present. To compare speeds across these fields, the figure uses the fixed scale $U_{c,s}$. The straining RMS speeds on this scale are $6.44\times10^{-4}$ and $8.50\times10^{-2}$ in the Taylor and Quincke states, respectively. The increase also remains substantial when each speed is nondimensionalised by its own $U_E$, giving $8.92\times10^{-4}$ and $5.03\times10^{-2}$. Mean rotation accounts for about $89\%$ of the squared total surface speed in the Quincke state. Subtracting it reveals the recovered straining circulation more clearly.

The recovered circulation coexists with a strongly nonuniform surfactant distribution. The concentration ranges are $0.411$--$1.555$ and $0.588$--$1.710$ in the Taylor and Quincke states, respectively. In the rotating state, electric and Marangoni tractions no longer nearly cancel, allowing substantial circulation despite the persistent concentration gradients. The spatial correlation coefficient of the mean-subtracted electric and Marangoni tractions changes from approximately $-1.000$ in the Taylor state to $-0.720$ in the Quincke state, indicating weaker anticorrelation of their spatial variations. The same Quincke concentration profile is shown in the total flow and its two velocity components in Fig.~\ref{fig:low_diffusion_flow}(b--d).

Figures~\ref{fig:surface_profiles} and \ref{fig:low_diffusion_flow} show that the rotating state has displaced charge and surfactant patterns together with an inclined straining circulation. The charge distribution, the surfactant distribution and the circulation need not have the same orientation. Their different evolution laws help explain why the electric and Marangoni stresses cease to cancel closely. Surface charge is supplied and relaxed by Ohmic currents, whereas insoluble surfactant is conserved. The electric traction is $\tau_E=qE_\theta$, with $E_\theta$ containing both the applied and induced fields, whereas the Marangoni traction is $\tau_M=-\Ma\,\partial_\theta\Gamma/(1-\chi\Gamma)$. A displacement of the charge relative to the applied field therefore need not change the electric traction in the same way as a displacement of surfactant changes the Marangoni traction. The orientation of the resulting circulation is quantified below.

\begin{figure}[tp]
 \centering
 \includegraphics[width=0.95\columnwidth]{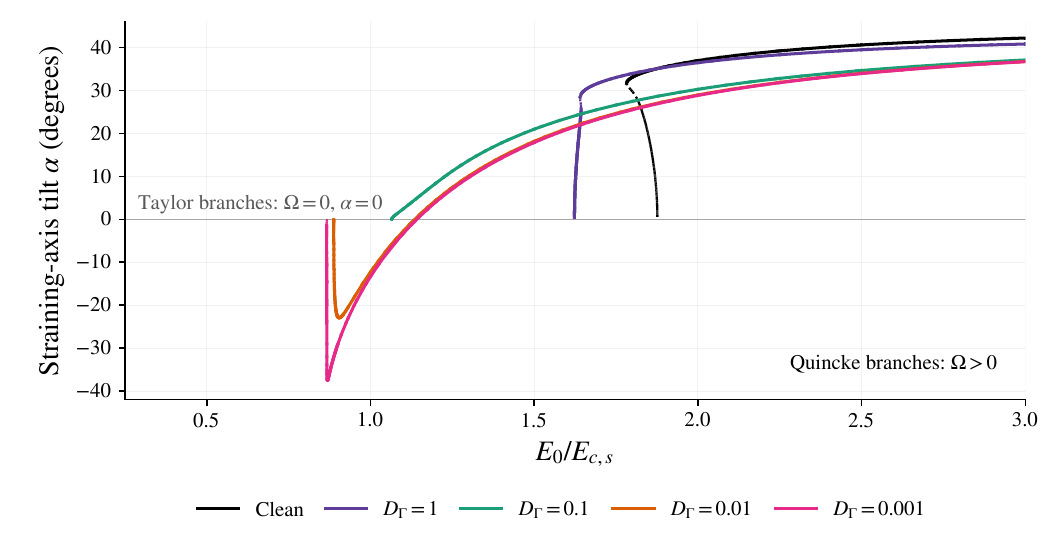}
 \caption{Central extensional-axis tilt $\alpha$ versus applied electric field for the clean drop and insoluble surfactant at the indicated diffusivities. Parameters are $R=S=2$, $M=1$, $B=1$ for the surfactant cases, $\chi=0.2$, $\Dq=10^{-3}$ and $\Ks=0$. Coloured curves show the branches with positive mean rotation, $\Omega>0$; the non-rotating Taylor solutions have $\alpha=0$. Solid portions denote stable states and dotted portions unstable or marginal states under the same imposed half-turn symmetries as Fig.~\ref{fig:rotation_branches}. The low-diffusion curves include additional near-onset samples. Their negative tilt occurs after the drop onset, which is below the solid reference field $E_{c,s}$. The angle is unchanged by the choice of velocity scale.}
 \label{fig:straining_tilt}
\end{figure}

Figure~\ref{fig:straining_tilt} quantifies the orientation using the extensional eigenvector of the rate-of-strain tensor at the drop centre. Writing the second harmonic of the straining surface velocity as
\[
 u_{{\rm str},2}=a_2\cos(2\theta)+b_2\sin(2\theta),
\]
the extensional-axis angle is defined, for $A_2=(a_2^2+b_2^2)^{1/2}>0$, by
\begin{equation}
 \cos(2\alpha)=\frac{b_2}{A_2},\qquad
 \sin(2\alpha)=-\frac{a_2}{A_2}.
 \label{eq:straining_tilt}
\end{equation}
The angle is measured from $+z$ towards $+x$ and is defined modulo $180^\circ$. When $A_2=0$, the central rate of strain vanishes and this orientation is undefined. Only the second surface-velocity harmonic contributes to the rate-of-strain tensor at the drop centre. Higher harmonics remain in the complete circulation and contribute to strain away from the centre. The Taylor solutions shown here have $\alpha=0$. For the clean drop and the surfactant cases with $\DG=1$ and $0.1$, the central extensional axis has positive tilt along the positive-rotation branches over the displayed fields. At lower diffusivities, $\DG=0.01$ and $0.001$, the tilt initially becomes negative after the respective onsets at $E_{c,d}/E_{c,s}=0.8876$ and $0.8663$. The sampled minima are approximately $-22.9^\circ$ and $-37.4^\circ$, respectively. The tilt subsequently crosses zero near $E_0/E_{c,s}=1.14$ and becomes positive. The mean rotation remains positive throughout: the reversal concerns the orientation of strain, rather than the direction of rotation. 

We also measure the area-averaged strain intensity in the interior disk. With the rate-of-strain tensor
\begin{equation}
 \bm S=\frac12\left[\grad\bu^-+(\grad\bu^-)^T\right],
 \label{eq:strain_tensor_results}
\end{equation}
we define
\begin{equation}
 \mathcal S=\frac{1}{A}\int_A \mathrm{Tr}(\bm S^2)\,dA.
 \label{eq:global_strain_results}
\end{equation}
Rigid rotation has zero strain, so $\mathcal S$ measures the straining part of the interior flow. In the $(z,x)$ coordinates, the rotation tensor associated with the mean rotational component is
\begin{equation}
 \bm W_\Omega=\frac12\left[\grad\bu_{\rm rot}-(\grad\bu_{\rm rot})^T\right]
 =\begin{pmatrix}0&-\Omega\\ \Omega&0\end{pmatrix}.
 \label{eq:mean_rotation_tensor}
\end{equation}
Its squared norm defines the mean-rotation intensity,
\begin{equation}
 \mathcal R=\bm W_\Omega:\bm W_\Omega=2\Omega^2.
 \label{eq:global_rotation_results}
\end{equation}
We then define
\begin{equation}
 \Theta=\frac{\mathcal S-\mathcal R}{\mathcal S+\mathcal R}.
 \label{eq:Theta_results}
\end{equation}
Thus $\Theta=1$ in the non-rotating Taylor regime and $\Theta=-1$ for pure rigid-body rotation; intermediate values compare the relative contributions of interior straining and mean rotation.

\begin{figure}[tp]
 \centering
 \includegraphics[width=0.95\columnwidth]{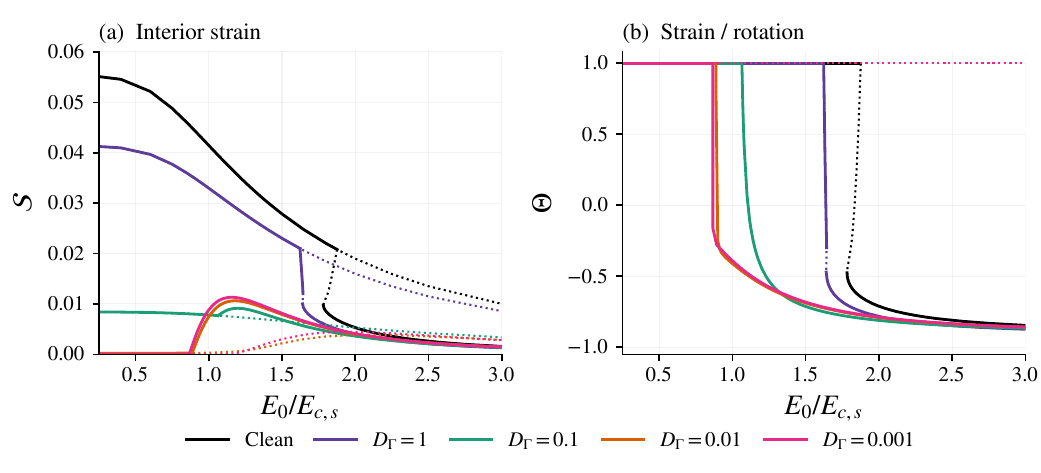}
 \caption{Measures of interior flow for $R=S=2$, $M=1$ and $B=1$. Left: area-averaged strain intensity $\mathcal S=\langle\mathrm{Tr}(\bm S^2)\rangle_A$. Right: $\Theta$ from Eq.~\eqref{eq:Theta_results}; the Taylor regime has $\Theta=1$, while negative values indicate that the mean-rotation intensity $\mathcal R$ exceeds the strain intensity.}
 \label{fig:global_strain}
\end{figure}

The same recovery is visible in these measures of the interior flow. Figure~\ref{fig:global_strain} shows the area-averaged strain intensity $\mathcal S$ and $\Theta$. For weakly diffusing surfactant the Taylor regime has very small strain, consistent with the nearly immobilised interface. Upon the onset of Quincke rotation, $\mathcal S$ rises from this suppressed value before decaying at larger fields in the electric velocity scaling. At the same time, $\Theta$ decreases from its Taylor value $+1$ and becomes negative. The rotating state is therefore dominated by mean rotation but still contains a finite straining circulation, and the relative importance of the two changes along the branch.

\subsection{Temporal evolution of the tangential stress balance}
\label{sec:stress_results}

\begin{figure}[tp]
 \centering
 \includegraphics[width=0.95\columnwidth]{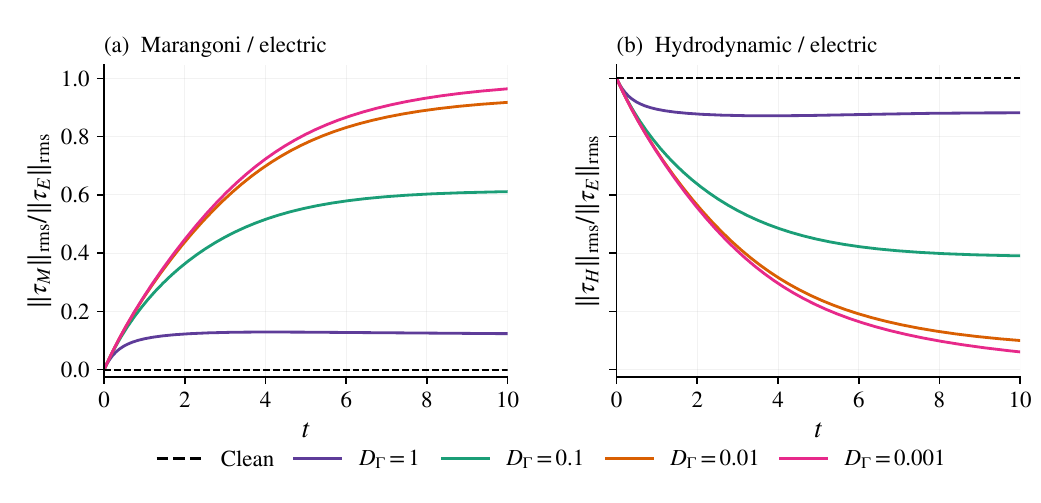}
 \caption{Temporal tangential stress balance for the fixed-circle Taylor state with $R=S=2$, $M=1$, $\ReE=B=1$, $\chi=0.2$, $\Dq=10^{-3}$ and $\Ks=0$. Left: $\|\tau_M\|_{\rm rms}/\|\tau_E\|_{\rm rms}$. Right: $\|\tau_H\|_{\rm rms}/\|\tau_E\|_{\rm rms}$. Colours denote surface diffusivity. Dashed black lines are the clean-interface values. Time is in units of $\tau^+$. The interface initially has $q=0$ and $\Gamma=1$.}
 \label{fig:stress_ratios}
\end{figure}

We follow the temporal evolution of the tangential stress balance in Fig.~\ref{fig:stress_ratios} using the baseline electrical parameters $R=S=2$, with $M=1$, $B=1$, $\ReE=1$, $\chi=0.2$, $\Dq=10^{-3}$ and $\Ks=0$. The field is applied at $t=0$ to an initially uncharged interface with uniform surfactant concentration. Since $RS>1$, the Taylor circulation is directed from the poles towards the equator.

The left panel gives the RMS Marangoni-to-electric traction ratio and the right panel the hydrodynamic-to-electric ratio. For a clean interface these ratios are zero and one, respectively. The ratios quantify how surfactant redistribution transfers the tangential stress balance from viscous to Marangoni stresses as the concentration gradients develop. These ratios of norms need not sum exactly to one because the local tractions have spatial structure and signs.

This behaviour is the finite-diffusion counterpart of the equilibrium immobilisation identified by \citet{Nganguia2013}. For a non-diffusing insoluble surfactant, they showed that surfactant redistribution can bring the interface to complete stagnation, so that the viscous tangential stress vanishes and the tangential electric stress is balanced solely by the Marangoni stress. The present calculations recover the approach to that stress balance as $\DG$ decreases. For any finite $\DG>0$, however, exact stagnation cannot persist in a steady state with a non-uniform surfactant distribution: in the absence of advection, surface diffusion would relax $\Gamma$ towards uniformity and eliminate the Marangoni stress. A small residual Taylor circulation therefore remains and continuously maintains the surfactant gradient.

Surfactant advection generates concentration gradients whose Marangoni stresses balance most of the tangential electric traction, leaving weak viscous stresses and a small circulation. Increasing surfactant diffusion or introducing sufficiently rapid exchange with the bulk weakens these gradients and restores interfacial motion. Quincke rotation provides a further route to remobilisation and recovery of straining circulation, as demonstrated in Figs.~\ref{fig:mobility} and \ref{fig:low_diffusion_flow}.

\subsection{How surfactant transport controls the onset of Quincke rotation}
\label{sec:onset_results}

\begin{figure}[tp]
 \centering
 \includegraphics[width=0.95\columnwidth]{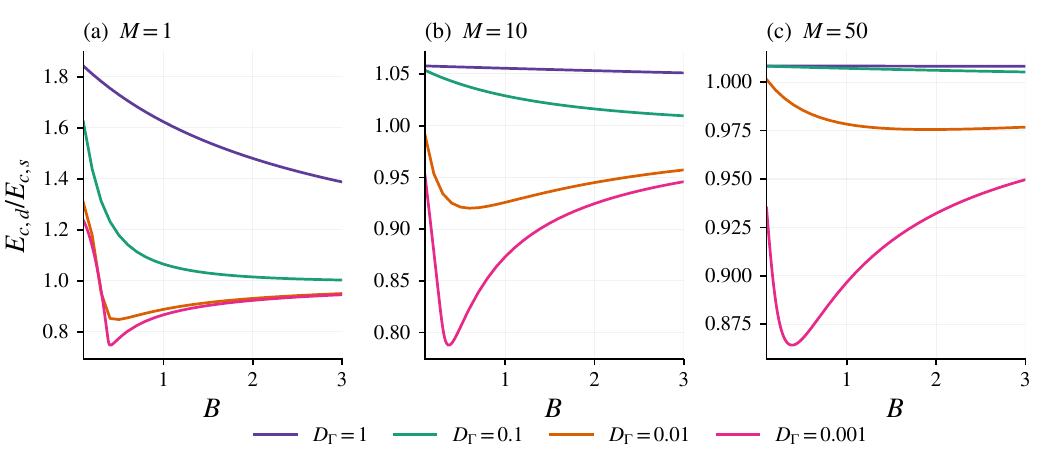}
 \caption{Critical field for Quincke rotation as a function of the field-independent surfactant strength $B=\ReE\Ma$. Panels show $M=1$, $10$ and $50$; colours denote $\DG$. The ordinate is $E_{c,d}/E_{c,s}$, where $E_{c,d}$ is the Taylor-branch linear-instability field and $E_{c,s}$ the common rigid-cylinder reference. Coverage is fixed at $\chi=0.2$.}
 \label{fig:onset_strength}
\end{figure}

The dependence of the critical field for Quincke rotation on surfactant strength is summarised in Fig.~\ref{fig:onset_strength}. The trends depend strongly on both $M$ and $\DG$. At $M=1$, large $\DG$ smooths the surfactant distribution and the critical field remains well above the rigid-cylinder value. Increasing $B$ nevertheless lowers the critical field. At smaller $\DG$ the reduction is much stronger, and rotation can begin below $E_{c,s}$. The lowest-diffusion curves are non-monotonic in $B$: the critical field reaches a minimum at intermediate surfactant strength and then rises again. A similar, though progressively weaker, non-monotonicity persists for $M=10$ and $50$. The effect of viscosity ratio is substantial; note that the y-axis range changes between the three panels of Fig.~\ref{fig:onset_strength}. The non-monotonic dependence shows that surfactant cannot be represented simply as progressively rigidifying the interface. Changing $B$ modifies both the Taylor-regime base state and the coupled charge--surfactant dynamics at the onset of Quincke rotation.

For $M=1$, the sampled minimum is $E_{c,d}/E_{c,s}=0.7472$ at $B=0.4$ for $\DG=10^{-3}$, and $0.8476$ at $B=0.5$ for $\DG=10^{-2}$. Throughout these sweeps $\chi=0.2$ is held fixed. Consequently, this non-monotonicity cannot be explained by simultaneous variation of $\Gamma_{\mathrm{eq}}$ in the numerator and denominator of the Langmuir stress law. Even at fixed $\chi$, changing $B$ changes the spatial base state and the linearised coupling that sets $E_{c,d}$. A small value of $B$ is also not necessarily a weak-feedback limit at small $\DG$: the leading immobilisation parameter scales as $B/\DG$, as derived below. 

\begin{figure}[tp]
 \centering
 \includegraphics[width=0.95\columnwidth]{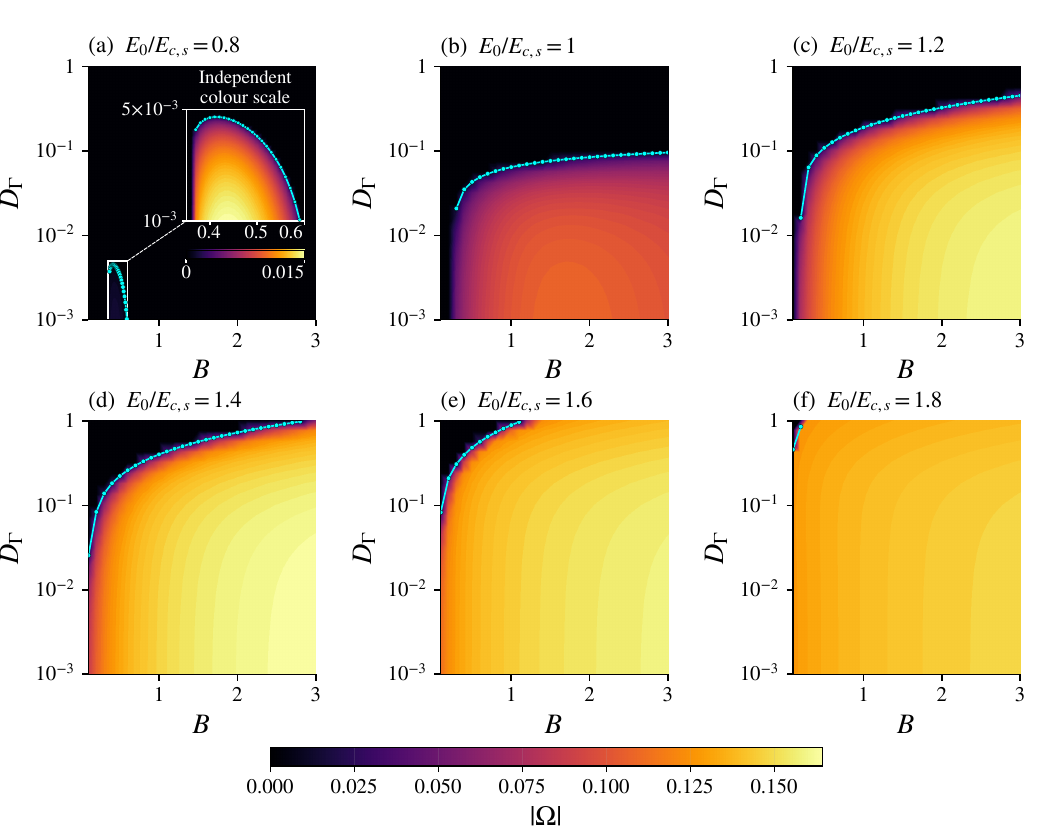}
 \caption{Rotation maps in the $(B,\DG)$ plane for $M=1$ at the indicated applied electric fields. Colour shows the dimensionless steady rotation magnitude $|\Omega|$ in the electroviscous time scaling; cyan symbols mark the onset of Quincke rotation for perturbations preserving the imposed half-turn symmetries. Black regions denote parameter combinations for which the selected solution is the stable non-rotating Taylor state. The inset magnifies the small rotating pocket at $E_0/E_{c,s}=0.8$ and uses an independent colour scale. At each parameter combination the map selects the Taylor state while it is stable, and otherwise a computed stable rotating state; it does not enumerate all coexisting attractors.}
 \label{fig:rotation_maps}
\end{figure}

The two-parameter organisation is made explicit in Fig.~\ref{fig:rotation_maps}, which maps the rotation magnitude in the $(B,\DG)$ plane at six fixed electric fields for $M=1$. The cyan points mark the onset of Quincke rotation. At $E_0/E_{c,s}=0.8$ the rotating region is confined to a very small pocket at low diffusion and intermediate $B$; the inset magnifies this pocket and shows that it is centred around $B\simeq0.4$ and $\DG$ of a few $10^{-3}$, consistent with the minimum in the low-diffusion onset curves of Fig.~\ref{fig:onset_strength}. By $E_0/E_{c,s}=1$ the pocket has expanded markedly, and for fields of $1.2$--$1.8$ the rotating state occupies an increasingly broad portion of the parameter plane. The onset boundary is strongly curved: neither increasing surfactant strength nor decreasing diffusion changes the onset monotonically over the whole parameter range. The boundary reflects the competition between surfactant advection and diffusion, coupled to the flow through Marangoni stresses.

\begin{figure}[tp]
 \centering
 \includegraphics[width=0.95\columnwidth]{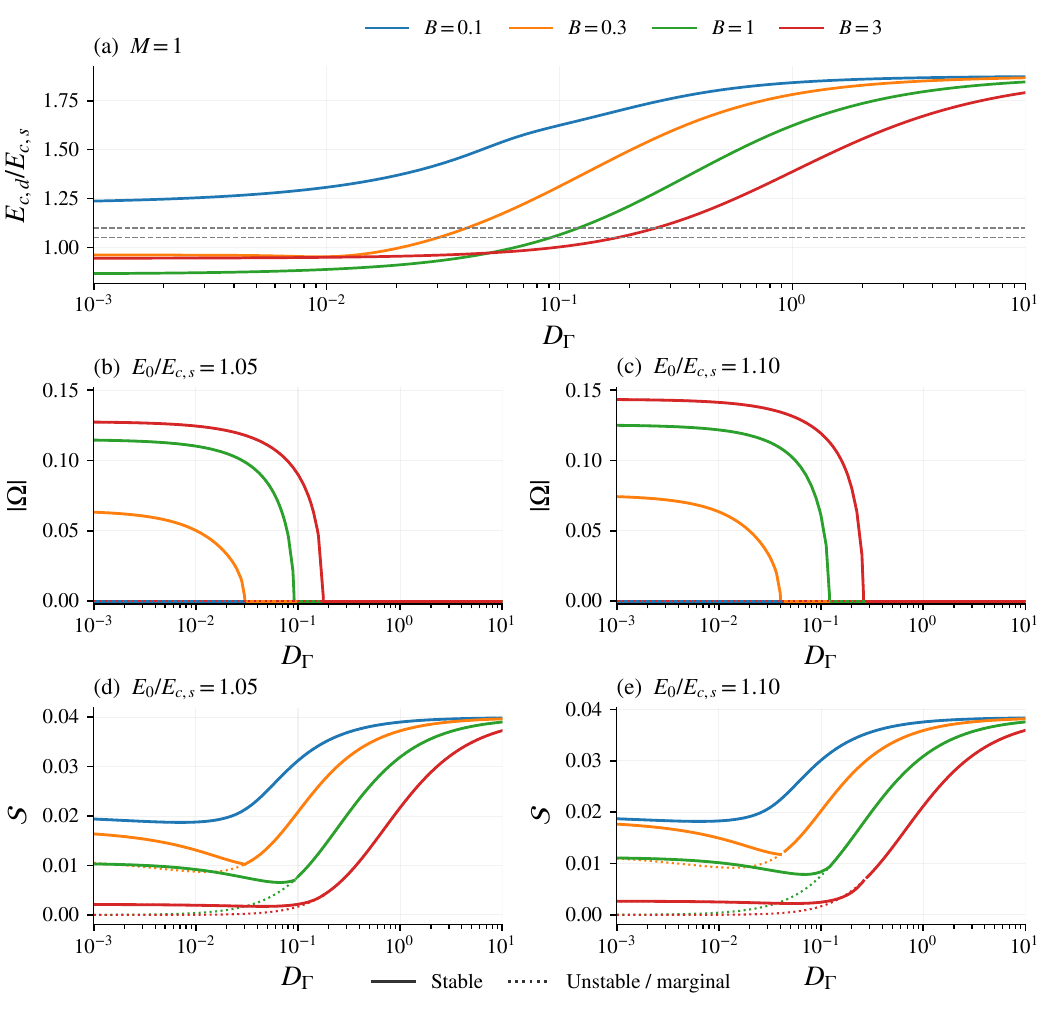}
 \caption{Diffusion cuts at $M=1$. Top: critical field for Quincke rotation versus $\DG$ for $B=0.1$, $0.3$, $1$ and $3$; the horizontal dashed levels correspond to the two fixed electric fields used below. Middle: steady rotation magnitude at $E_0/E_{c,s}=1.05$ and $1.1$. Bottom: corresponding area-averaged strain intensity. Solid curves denote states classified as stable to perturbations preserving the imposed half-turn symmetries, and dotted portions denote unstable or marginal continuations.}
 \label{fig:diffusion_cuts}
\end{figure}

Figure~\ref{fig:diffusion_cuts} takes cuts through the same parameter space along the $\DG$ direction at fixed $B$. The upper panel plots the critical field for Quincke rotation against $\DG$ for four surfactant strengths. At large $\DG$ all curves approach the clean-drop value because diffusion prevents appreciable surfactant gradients. Decreasing $\DG$ then separates the families: weak surfactant ($B=0.1$) still delays the onset relative to the rigid-cylinder value, whereas $B=0.3$--$3$ can drive it below $E_{c,s}$. The two middle panels show the finite rotation present at fixed electric fields $E_0/E_{c,s}=1.05$ and $1.1$. Remarkably, $\DG$ itself acts as a bifurcation parameter: at fixed electric field and surfactant strength $B$, reducing surfactant diffusion can trigger Quincke rotation. The lower panels show the corresponding area-averaged strain intensity. A state can therefore rotate strongly while its strain intensity remains small: rotation can begin before the Taylor circulation is fully restored.

\subsection{Soluble exchange and the hidden-variable problem}
\label{sec:solubility_results}

All results above use an insoluble monolayer, with surface diffusion as its only explicit surfactant-relaxation process. The interpretation of surface-speed measurements nevertheless extends beyond this particular transport model. In the hidden-variable perspective of \citet{ManikantanSquires2020}, the same macroscopic response can arise from different combinations of elasticity, transport and surface rheology. The translating-disk analysis of \citet{Elfring2016}, for example, shows how Marangoni stresses can complicate the interpretation of interfacial drag as a surface-viscosity measurement. \citet{LiManikantan2024} demonstrated this ambiguity in the stability of liquid jets: when exchange is rapid compared with instability growth and surface diffusion is negligible, Marangoni stresses coupled to finite adsorption and desorption contribute to the dispersion relation in the same form as an intrinsic surface viscosity. They also showed that faster exchange and stronger surface diffusion both relax concentration gradients and weaken the Marangoni resistance to interfacial deformation. The present model has no intrinsic surface viscosity, and so tests diffusion versus exchange.

Dip coating provides useful examples of both the ambiguity and ways to constrain it. \citet{Scheid2010} attributed residual film thickening under conditions of rapid surfactant supply to surface rheology. A thickness change is therefore not automatically evidence for a persistent Marangoni gradient. \citet{Delacotte2012} varied withdrawal speed, concentration and surfactant chemistry, and connected changes in thickening to confinement of the surfactant supply and, for some systems, adsorption barriers. Neither finite-film reservoir depletion nor intrinsic surface viscosity is included in our local-exchange model. Conversely, \citet{Champougny2015} compared free-film drawing and dip coating using a model in which surfactant exchange is negligible over the flow timescale, and obtained a consistent surface elasticity that also agreed with independent Langmuir-trough measurements. Their comparison illustrates how different forcing geometries and independent measurements of surface elasticity in a Langmuir trough can constrain an interpretation that film thickness alone leaves ambiguous.

\begin{figure}[t]
 \centering
 \includegraphics[width=0.65\columnwidth]{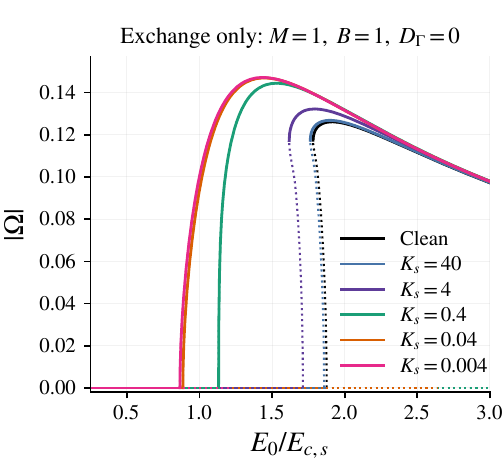}
 \caption{Steady rotation for the exchange-only family $M=1$, $B=1$, $\DG=0$. The clean interface is compared with $\Ks=40$, $4$, $0.4$, $0.04$ and $0.004$. Solid lines denote states classified as stable to perturbations preserving the imposed half-turn symmetries; dotted lines denote unstable or marginal continuations.}
 \label{fig:solubility_rotation}
\end{figure}

Here the imposed electric field changes the coupled charge and surfactant distributions and can drive a symmetry-breaking rotational instability. This offers a complementary, spatially resolved setting for testing whether similar surface speeds and rotation thresholds conceal distinct surfactant transport mechanisms. To examine whether exchange produces changes similar to those caused by surface diffusion, we set $\DG=0$ and allow exchange with a well-mixed reservoir. Figure~\ref{fig:solubility_rotation} shows the resulting branches for $M=1$ and $B=1$. When exchange is rapid ($\Ks=40$ and $4$), concentration variations are continually relaxed and the rotating branch remains close to the clean-drop response, including a subcritical unstable segment connected to the stable rotating branch at a saddle-node bifurcation. As exchange is slowed, the critical field for Quincke rotation shifts to lower values and the onset becomes supercritical, as also found for weak surface diffusion in Fig.~\ref{fig:rotation_branches}. Movies~S5 and S6 illustrate the development of Taylor circulation and Quincke rotation with rapid exchange ($\Ks=4$) and no surface diffusion; their parameters are listed in the Supplementary Material.

\begin{figure}[tp]
  \centering
  \includegraphics[width=0.95\columnwidth]{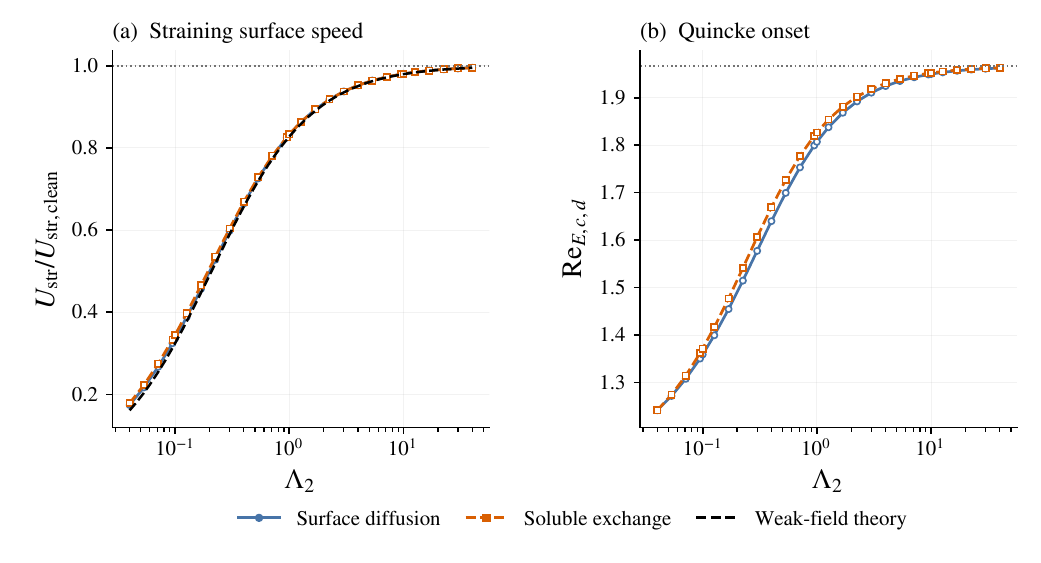}
  \caption{Matched diffusion and soluble-exchange relaxation for $M=5$, $B=1$, $R=S=2$ and $\Dq=10^{-3}$. Left: straining surface speed relative to the clean interface at the same Taylor-regime field, $\ReE=1.0881$ ($E_0/E_{c,s}\simeq0.85$). Right: critical electric Reynolds number $\mathrm{Re}_{E,c,d}$. Diffusion-only cases have $\Ks=0$ and exchange-only cases have $\DG=0$; the horizontal coordinate is the common quadrupolar relaxation rate $\Lambda_2$. The dashed black curve in (a) is the weak-field prediction of Eq.~\eqref{eq:immobilisation_number}, with no fitted parameters.}
  \label{fig:solubility_collapse}
\end{figure}

\begin{figure}[tp]
 \centering
 \includegraphics[width=0.95\columnwidth]{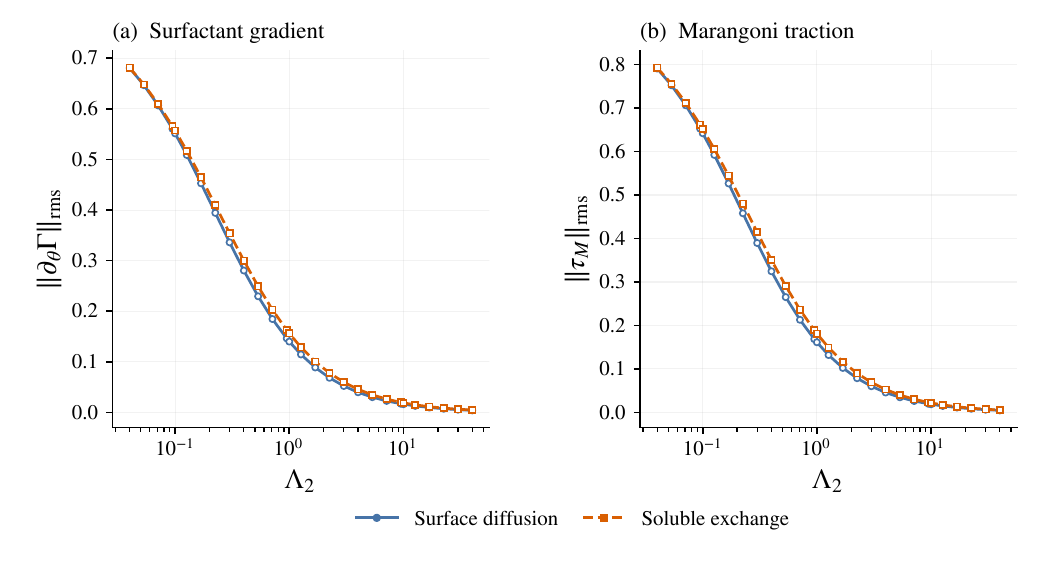}
 \caption{Mechanism underlying the collapse in Fig.~\ref{fig:solubility_collapse}: RMS surfactant gradient (left) and RMS Marangoni traction (right) for matched diffusion-only and exchange-only calculations.}
 \label{fig:solubility_mechanism}
\end{figure}

The correspondence between diffusion and exchange is nevertheless incomplete. In particular, the subcritical branch at $\Ks=4$ differs from the $\DG=1$ case, where rotation emerges supercritically and a much narrower bistable interval develops at finite rotation amplitude. The paired values $\Ks=40$, $4$, $0.4$, $0.04$ and $0.004$ and $\DG=10$, $1$, $0.1$, $0.01$ and $0.001$, respectively, match the relaxation rate of the quadrupolar surfactant mode, as derived below. This matching gives the same leading-order response in the weak-field Taylor regime, but does not imply identical nonlinear bifurcation diagrams. Charge and surfactant transport are nonlinearly coupled through the interfacial flow, and diffusion and exchange relax the higher spatial harmonics at different rates. Increasing the viscosity ratio weakens the straining circulation and its feedback on the interfacial distributions, making the leading-order quadrupolar description more appropriate. We therefore use $M=5$ for the subsequent comparisons in Figs.~\ref{fig:solubility_collapse} and \ref{fig:solubility_mechanism}, where the two transport mechanisms produce closely similar responses. 

To establish the leading-order correspondence, we first linearise the surfactant equation about the uniform state. The diffusion and local-exchange operators damp a Fourier perturbation $\delta\Gamma_n e^{in\theta}$ at the rate due to diffusion or exchange alone,
\begin{equation}
 \Lambda_n=
 \begin{cases}
  n^2\DG, & \text{diffusion only }(\Ks=0),\\[1mm]
  \Ks, & \text{exchange only }(\DG=0).
 \end{cases}
 \label{eq:Lambda_n}
\end{equation}
The comparison below keeps only one of these two transport mechanisms active. The Taylor circulation is dominated by the quadrupolar surfactant mode, so $\Lambda_2=4\DG$ for the diffusion-only family and $\Lambda_2=\Ks$ for the exchange-only family.

The collapse expected from $\Lambda_2$ can be made explicit in the weak-field Taylor regime. Retaining only the dominant quadrupolar mode, write
\begin{equation}
 u_\theta=U_2\sin2\theta,
 \qquad
 \Gamma=1+G_2\cos2\theta.
 \label{eq:weak_mode_ansatz}
\end{equation}
For small concentration variations about $\Gamma=1$, we linearise the surfactant transport equation and the Langmuir equation of
state while retaining finite $\chi$. The steady transport equation then gives
\begin{equation}
 G_2=-\frac{2\ReE}{\Lambda_2}U_2.
 \label{eq:G2_linear}
\end{equation}
The Marangoni traction opposes the surface motion, with amplitude $-4BU_2/[(1-\chi)\Lambda_2]$, while the hydrodynamic traction jump has amplitude $-4(1+M)U_2$. At leading order in weak electric forcing, the electric traction is unchanged by surfactant. Comparing with the clean Taylor solution at the same field therefore gives
\begin{equation}
 \frac{U_2}{U_{2,\mathrm{clean}}}
 =\left(1+\mathcal N_{\rm imm}\right)^{-1},
 \qquad
 \mathcal N_{\rm imm}\equiv
 \frac{B}{(1+M)(1-\chi)\Lambda_2}.
 \label{eq:immobilisation_number}
\end{equation}
We define $\mathcal N_{\rm imm}$ as a dimensionless immobilisation parameter for this weak-field quadrupolar approximation. It measures the strength of Marangoni suppression relative to the clean Taylor circulation: $\mathcal N_{\rm imm}\ll1$ gives weak suppression, whereas $\mathcal N_{\rm imm}\gg1$ gives strong suppression. Equation~\eqref{eq:immobilisation_number} predicts the same leading-order surface-speed reduction for diffusion and exchange when their quadrupolar relaxation rates $\Lambda_2$ are equal. The derivation is given in the Supplementary Material.
By matched relaxation we mean choosing paired diffusion-only and exchange-only cases with the same $\Lambda_2$, equivalently $\Ks=4\DG$ between the paired cases. These calculations use $M=5$, $R=S=2$, $\Dq=10^{-3}$, $B=1$ and $\chi=0.2$. The Taylor-regime speeds, concentration gradients and stresses are compared at the same pre-onset field $\ReE=1.0881$, corresponding to $E_0/E_{c,s}\simeq0.85$; the critical fields are determined separately by varying the applied field for each parameter set. The larger viscosity ratio is used here to compare the two surfactant-relaxation mechanisms away from the sharpest charge-density gradients at low $M$.

The two surfactant-transport mechanisms give almost the same straining surface speed and critical field for Quincke rotation in Fig.~\ref{fig:solubility_collapse}. The straining surface speed, normalised by its clean value at the same field, nearly collapses for diffusion and exchange over more than three decades in $\Lambda_2$. The weak-field calculation in Eq.~\eqref{eq:immobilisation_number} shows that this agreement is expected for the dominant quadrupolar mode: at leading order the surface-speed ratio depends on the surfactant-transport parameters only through $\Lambda_2$. The corresponding critical fields also lie close together and approach the clean-drop value as $\Lambda_2$ increases. The nonlinear calculations show that the two mechanisms produce similar surface speeds and rotation thresholds even at finite field, while their surfactant distributions differ.

The associated surfactant gradients and Marangoni stresses are shown in Fig.~\ref{fig:solubility_mechanism}. Both $\|\partial_\theta \Gamma\|_{\rm rms}$ and $\|\tau_M\|_{\rm rms}$ decrease with $\Lambda_2$, following broadly similar trends for diffusion and exchange. At small $\Lambda_2$, the Taylor circulation produces large concentration gradients and a strong Marangoni stress. Either faster surface diffusion or faster exchange with the reservoir reduces those gradients, weakens the Marangoni stress and remobilises the interface. At the level of the dominant mode, it matters little whether the decay is produced by surface diffusion or by local exchange with the bulk reservoir.

\begin{figure}[t]
 \centering
 \includegraphics[width=0.65\columnwidth]{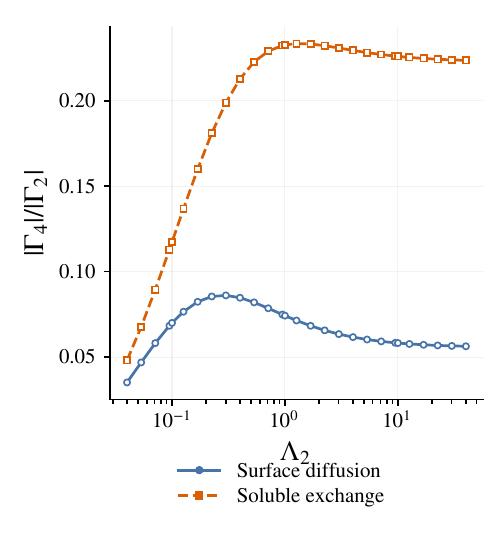}
 \caption{Higher-harmonic measure for the matched-relaxation comparison. The ratio $|\Gamma_4|/|\Gamma_2|$ separates surface diffusion and soluble exchange even when their straining surface speed and critical field for Quincke rotation nearly collapse at the same $\Lambda_2$.}
 \label{fig:higher_harmonics}
\end{figure}

The surfactant distributions are not the same, however. Surface diffusion damps mode $n$ at rate $n^2\DG$, whereas local exchange with the well-mixed reservoir damps every mode at $\Ks$. At equal $\Lambda_2$, the diffusion and exchange calculations give nearly identical straining surface speeds and critical fields, but different amplitudes of the higher Fourier modes of surfactant concentration. At $\Lambda_2=1$, for example, the computed Taylor surface-speed ratios are $0.8326$ and $0.8332$ for diffusion and exchange, respectively, while $|\Gamma_4|/|\Gamma_2|$ is $0.0743$ and $0.2326$, a difference of more than a factor of three (Fig.~\ref{fig:higher_harmonics}). Surface speed and rotation threshold alone may therefore be insufficient to distinguish diffusion from exchange. The angular variation of surfactant concentration provides an additional measure that distinguishes the two mechanisms in these calculations.

The distinction arises because matching the relaxation of the $n=2$ mode does not match that of the higher modes. For example, setting $\Ks=4\DG$ makes the two rates equal for $n=2$, but diffusion damps $n=4$ four times faster than the paired exchange process. The ratio $|\Gamma_4|/|\Gamma_2|$ therefore distinguishes the calculated concentration distributions even when their surface speeds are very similar. Resolved concentration measurements could provide this additional information. The prescribed reservoir assumes that bulk transport maintains a spatially uniform concentration adjacent to the interface. Resolving finite bulk transport would relax this assumption and could modify the comparison between exchange and surface diffusion.

\section{Conclusions}

In this paper we studied a surfactant-covered leaky-dielectric drop in a uniform dc electric field using a fixed circular interface. The model retains the full nonlinear surface-charge convection together with surfactant advection, surface diffusion, Marangoni stresses and, separately, local exchange with a well-mixed bulk reservoir. This setting resolves the charge and surfactant fields around the entire interface without a small-$\ReE$ expansion or a truncation to the first few harmonics.

The problem couples two advected interfacial fields. Surface charge sets the electric shear stress and is convected by the flow, while the same flow redistributes surfactant and generates a Marangoni stress. In the non-rotating Taylor regime, our results recover the Marangoni immobilisation mechanism previously identified for a non-diffusing surfactant: as surface diffusion decreases, the Marangoni stress carries most of the tangential electric-stress balance and the hydrodynamic stress and Taylor circulation become small. With finite surface diffusion the interface does not become exactly stagnant, because diffusion continually smooths the surfactant gradient, but it can become nearly immobilised. Retaining charge convection at arbitrary strength resolves the sharp charge-density gradients that develop in the Taylor regime and determine the base state from which Quincke rotation begins.

When charge convection destabilises the nearly immobilised Taylor state, Quincke rotation restores both interfacial motion and a substantial straining circulation. At weak surfactant diffusion, this recovery occurs despite persistent concentration gradients: electric and Marangoni tractions no longer nearly cancel. The circulation also changes orientation. For the clean drop and larger surfactant diffusivities, its central extensional axis has positive tilt on the positive-rotation branches. At low diffusivity, the tilt changes from negative to positive as the field increases, while the direction of rotation remains unchanged.

Surfactant also changes the Quincke rotational instability itself. It can lower the critical field below the rigid-cylinder value, produce a non-monotonic dependence of the critical field on surfactant strength, and change the onset from subcritical to supercritical. At fixed electric field and surfactant strength, the surface diffusivity can itself act as a bifurcation parameter, with decreasing $\DG$ driving the drop from the Taylor regime into Quincke rotation. The Taylor-regime base state and the rotational instability are therefore coupled through the redistributed surfactant and charge fields.

Surface diffusion and exchange with a well-mixed bulk reservoir act through different physical processes. In the $M=5$ comparisons, the straining surface speed and critical field for Quincke rotation nearly coincide when the dominant quadrupolar surfactant mode has the same relaxation rate in the two systems. A weak-field calculation predicts this agreement in surface speed. The nonlinear calculations retain higher spatial modes, whose amplitudes differ because diffusion and exchange damp them at different rates. Surface speed and rotation threshold alone can therefore conceal the transport mechanism, whereas the spatial variation of surfactant concentration supplies additional information.

The circular geometry suppresses shape changes but allows charge convection at arbitrary strength and resolves the sharp interfacial charge and surfactant gradients that make fully resolved three-dimensional continuation expensive, particularly at low viscosity ratios. The results are therefore most directly applicable when surface tension is sufficiently strong that deformation remains small. Extending the present diffusion--exchange comparison and rotation-induced remobilisation mechanism to a fully resolved deforming three-dimensional interface, or to a finite bulk surfactant reservoir, would require substantially greater computational effort.

Another important extension is to determine how interactions between drops modify the immobilisation and rotation-induced remobilisation identified here. Surfactant can weaken electrohydrodynamic interactions and increase the relative importance of dielectrophoresis when charge convection is negligible.\citep{SorgentoneVlahovska2021} The competition between these interactions also governs drop trajectories and collective organisation in emulsions.\citep{KachWalkerKhair2022,KachWalkerKhair2023} More recently, a reduced model retaining charge convection has captured interacting clean drops undergoing Quincke rotation in the regime of large separation, high viscosity ratio and small deformation.\citep{McDougallInteractions2026} These studies raise the question of how surfactant modifies interactions when charge convection by the straining flow becomes appreciable, including whether rotation restores interactions weakened by Marangoni stresses. Addressing this question with spatially resolved nonlinear charge and surfactant transport would be computationally demanding, requiring sharp interfacial gradients to be resolved on multiple moving and deforming drops together with their coupled electric and hydrodynamic interactions.

\section*{Conflicts of interest}

There are no conflicts to declare.

\section*{Data availability}

Supplementary Material is available. Raw numerical data are available from the corresponding author upon reasonable request.

\section*{Acknowledgements}

This work was supported by the Leverhulme Trust
(Research Project Grant RPG-2026-185).

We thank Harishankar Manikantan for helpful discussions and valuable insights into the physical interpretation of the problem and Alexander Wray for constructive comments on the manuscript.

\clearpage
\newgeometry{margin=2cm}
\onecolumn
\makeatletter
\renewcommand{\normalsize}{\@setfontsize\normalsize{11pt}{13.6pt}%
  \abovedisplayskip 11pt plus3pt minus6pt
  \abovedisplayshortskip 0pt plus3pt
  \belowdisplayshortskip 6.5pt plus3.5pt minus3pt
  \belowdisplayskip\abovedisplayskip}
\renewcommand{\small}{\@setfontsize\small{10pt}{12pt}}
\renewcommand{\large}{\@setfontsize\large{12pt}{14pt}}
\renewcommand{\Large}{\@setfontsize\Large{14.4pt}{18pt}}
\makeatother
\normalsize
\sectionfont{\Large\bfseries}
\subsectionfont{\large\bfseries}
\captionsetup{format=plain,justification=justified,singlelinecheck=true,
              font=small,labelfont=bf,labelsep=space}
\setlength{\bibsep}{3pt}
\setlength{\parindent}{17pt}
\setcounter{section}{0}
\setcounter{subsection}{0}
\setcounter{equation}{0}
\setcounter{figure}{0}
\setcounter{table}{0}
\setcounter{page}{1}
\renewcommand{\thesection}{S\arabic{section}}
\renewcommand{\theequation}{S\arabic{equation}}
\renewcommand{\thefigure}{S\arabic{figure}}
\renewcommand{\thetable}{S\arabic{table}}
\renewcommand{\thepage}{S\arabic{page}}
\renewcommand{\theHsection}{supp.\arabic{section}}
\renewcommand{\theHsubsection}{supp.\arabic{section}.\arabic{subsection}}
\renewcommand{\theHequation}{supp.\arabic{equation}}
\renewcommand{\theHfigure}{supp.\arabic{figure}}
\renewcommand{\theHtable}{supp.\arabic{table}}
\renewcommand{\jump}[1]{\left[\!\left[#1\right]\!\right]}

\begin{center}
{\Large\bfseries Supplementary Material\par}
\vspace{0.4cm}
{\large\bfseries Immobilisation and remobilisation of surfactant-covered drop interfaces in electric fields: Simulations\par}
\vspace{0.4cm}
{\large Debasish Das\par}
\vspace{0.2cm}
{\small Department of Mathematics and Statistics, University of Strathclyde,\par
Livingstone Tower, 26 Richmond Street, Glasgow G1 1XH, United Kingdom.\par
E-mail: \href{mailto:debasish.das@strath.ac.uk}{debasish.das@strath.ac.uk}\par}
\end{center}

\section{Numerical method}
\label{app:numerics}

Notation and nondimensionalisation in this section are as in the main text.  For a circular interface, the bulk Laplace and Stokes problems can be solved analytically to express the interfacial electric field and velocity in terms of the surface charge and surfactant concentration. The resulting boundary-integral formulation uses linear operators that are diagonal in Fourier space and can be expressed using the periodic Hilbert transform. We follow the interfacial formulation used for the clean circular-drop problem by \citet{sm-PengSchnitzer2025}, express it in terms of $R$, $S$, $M$ and $\ReE$, and include surfactant transport and Marangoni stresses.

\subsection{Boundary operators and interfacial evolution equations}

Let $\mathcal H$ denote the circular Hilbert transform,
\begin{equation}
 \mathcal H[f](\theta)=\frac{1}{2\pi}\,\mathrm{p.v.}
 \int_{-\pi}^{\pi}f(\varphi)
 \cot\!\left(\frac{\theta-\varphi}{2}\right)d\varphi,
 \label{eq:app_hilbert}
\end{equation}
with the convention $\mathcal H[\cos(n\theta)]=\sin(n\theta)$. Eliminating the harmonic electric potentials gives the tangential electric field at the interface as
\begin{equation}
 E_\theta
 =\frac{\mathcal H[q]-2\sin\theta}{1+S}.
 \label{eq:app_E_surface}
\end{equation}
Similarly, eliminating the two Stokes fields gives a relation between the derivative of the surface velocity and the total non-hydrodynamic tangential traction,
\begin{equation}
 \frac{\partial u_\theta}{\partial\theta}
 =-\frac{1}{2(1+M)}
 \mathcal H\!\left[qE_\theta+\tau_M\right],
 \label{eq:app_v_surface}
\end{equation}
where
\begin{equation}
 \tau_M
 =-\Ma\frac{1}{1-\chi\Gamma}
 \frac{\partial\Gamma}{\partial\theta}.
 \label{eq:app_marangoni_surface}
\end{equation}
The mean surface velocity is fixed by the electric--hydrodynamic torque balance,
\begin{equation}
 \Omega=\langle u_\theta\rangle
 =-\frac{1}{1+S}\langle q\sin\theta\rangle.
 \label{eq:app_mean_v}
\end{equation}
Equations~\eqref{eq:app_E_surface}--\eqref{eq:app_mean_v} therefore reconstruct the electric shear and interfacial velocity directly from $q(\theta,t)$ and $\Gamma(\theta,t)$.

Substituting the interfacial normal electric fields into the charge-conservation equation gives
\begin{equation}
 \frac{\partial q}{\partial t}
 +\ReE\frac{\partial}{\partial\theta}(qu_\theta)
 +\frac{R+1}{R(1+S)}\,q
 =
 \Dq\frac{\partial^2 q}{\partial\theta^2}
 -\frac{2(RS-1)}{R(1+S)}\cos\theta.
 \label{eq:app_q_evolution}
\end{equation}
The forcing term changes sign with $RS-1=\tau^-/\tau^+-1$, which is also the material distinction relevant to the direction of the Taylor circulation. The surfactant equation is
\begin{equation}
 \frac{\partial\Gamma}{\partial t}
 +\ReE\frac{\partial}{\partial\theta}(\Gamma u_\theta)
 =
 \DG\frac{\partial^2\Gamma}{\partial\theta^2}
 +\Ks(1-\Gamma).
 \label{eq:app_G_evolution}
\end{equation}
The steady problem is therefore obtained by setting $\partial_t q=\partial_t\Gamma=0$ in Eqs.~\eqref{eq:app_q_evolution} and \eqref{eq:app_G_evolution}, with $E_\theta$, $\tau_M$ and $u_\theta$ determined from Eqs.~\eqref{eq:app_E_surface}--\eqref{eq:app_mean_v}. The clean-drop formulation is recovered by suppressing the Marangoni term and omitting Eq.~\eqref{eq:app_G_evolution}.

\subsection{Periodic discretisation and reconstruction}

We discretise $0\leq\theta<2\pi$ on an even periodic grid
\begin{equation}
 \theta_j=\frac{2\pi j}{N},
 \qquad j=0,\ldots,N-1,
 \qquad h=\frac{2\pi}{N}.
 \label{eq:app_grid}
\end{equation}
On a uniform grid the circular Hilbert transform is represented by its exact trigonometric collocation matrix,
\begin{equation}
 H_{jk}=
 \begin{cases}
 \displaystyle
 \frac{2}{N}\cot\!\left[\frac{\theta_j-\theta_k}{2}\right],
 &j-k\ {\rm odd},\\[2mm]
 0,&j-k\ {\rm even},
 \end{cases}
 \label{eq:app_Hmatrix}
\end{equation}
while first and second derivatives in the transport equations are evaluated with periodic centred differences,
\begin{equation}
 (D_1f)_j=\frac{f_{j+1}-f_{j-1}}{2h},
 \qquad
 (D_2f)_j=\frac{f_{j+1}-2f_j+f_{j-1}}{h^2}.
 \label{eq:app_derivatives}
\end{equation}
Indices are periodic. We combine centred differences for conservative interfacial transport with Fourier evaluation of the circular electric and hydrodynamic boundary operators on a uniform periodic grid. The Hilbert operator is spectrally exact on the retained trigonometric modes, while the transport discretisation determines the overall second-order spatial accuracy for smooth solutions. The same spatial discretisation is used for steady continuation, transient integration and linear stability. Spatial accuracy is assessed by successive grid refinement of the interfacial fields and their gradients.

In the implementation, the Hilbert transform is evaluated by its Fourier multiplier $-i\,\operatorname{sgn}(n)$, with the mean and Nyquist coefficients set to zero. The Nyquist mode is the highest grid wavenumber, $n=N/2$. The Marangoni traction is evaluated by applying the discrete derivative directly to the Langmuir expression for
$(\gamma-1)/\CaE$,
\begin{equation}
 (\tau_M)_j=D_1\!\left[\frac{\Ma}{\chi}
 \ln\!\left(\frac{1-\chi\Gamma}{1-\chi}\right)\right]_j.
 \label{eq:app_discrete_marangoni}
\end{equation}
This expression has zero discrete mean on the periodic grid and converges to Eq.~\eqref{eq:app_marangoni_surface}. At finite spacing it differs from applying a pointwise chain rule to $D_1\Gamma$. Its analytic variation, $D_1[-\Ma\,\delta\Gamma/(1-\chi\Gamma)]$, is used in the discrete Jacobian. The velocity is reconstructed directly in Fourier space. For each Fourier mode with $0<|n|<N/2$, the mobility multiplier is $[2(1+M)|n|]^{-1}$ acting on $\tau_E+\tau_M$; the mean is supplied by Eq.~\eqref{eq:app_mean_v}. This is equivalent to spectral integration of Eq.~\eqref{eq:app_v_surface}. The Nyquist velocity coefficient is set to zero. The steady residuals are
\begin{align}
 F_q={}&\ReE D_1(qu_\theta)
 +\frac{R+1}{R(1+S)}\,q
 -\Dq D_2q
 +\frac{2(RS-1)}{R(1+S)}\cos\theta,
 \label{eq:app_Fq}\\
 F_\Gamma={}&\ReE D_1(\Gamma u_\theta)
 -\DG D_2\Gamma+\Ks(\Gamma-1).
 \label{eq:app_FG}
\end{align}
The local exchange term therefore adds no new nonlocal operator: it enters directly in the surfactant residual and Jacobian. The steady Taylor- and Quincke-regime solutions considered here obey the half-turn symmetries
\begin{equation}
 q(\theta+\pi)=-q(\theta),
 \qquad
 \Gamma(\theta+\pi)=\Gamma(\theta).
 \label{eq:app_symmetry}
\end{equation}
These symmetries are imposed in the nonlinear solve. For an $N$-point grid only $N/2$ charge values are independent. For $\Ks=0$, the prescribed mean $\langle\Gamma\rangle=1$ removes the conserved surfactant-mass mode; for $\Ks>0$ the same mean is fixed by the reservoir equilibrium.

The branch and stability calculations use uniform grids with $N=128$, $256$ and $512$, with refined clean and exchange-only
branches using $N=1024$ and $2048$. Selected states on the clean and $\Ks=40$ branches are additionally corrected at $N=4096$
where required, using the same acceptance tolerances. Selected interfacial profiles were also recomputed on finer grids. Both low-diffusion states in Fig.~6 of the main text use $N=512$, whereas the clean Taylor states at $E_0/E_{c,s}=1.30$ and $1.80$ in Fig.~2 use $N=4096$. Narrow charge layers near convergent stagnation points make convergence particularly demanding at low viscosity ratio and weak charge diffusion.

\subsection{Newton iteration and pseudo-arclength continuation}

The non-rotating Taylor branch is initiated from the weak-field solution, for which $q\propto\cos\theta$ and $\Gamma=1$, and continued in $\ReE$ using the converged solution at the previous parameter value as the initial guess. At fixed $\ReE$ the nonlinear system $\bm F(\bm x,\ReE)=\bm0$ is solved by Newton iteration with an analytically differentiated discrete Jacobian. A backtracking line search is applied whenever the full Newton step does not reduce the residual. Continuation using $\ReE$ as the parameter cannot pass a fold, where the steady solution ceases to be a single-valued function of $\ReE$. Rotating branches are therefore followed by pseudo-arclength continuation in the state and logarithmic field parameter $\rho=\ln\ReE$. With $n_x$ independent real state coordinates, the metric is
\begin{equation}
 \| (\delta\bm x,\delta\rho)\|_W^2
 =\frac{1}{n_x}\|\delta\bm x\|^2+(\delta\rho)^2.
 \label{eq:app_palc_metric}
\end{equation}
Two converged rotating states define a unit secant tangent $(\bm t_x,t_\rho)$ in this metric. A predictor $(\bm x_p,\rho_p)$ is corrected by solving
\begin{align}
 \bm F(\bm x,e^\rho)&=\bm0,
 \label{eq:app_palc_F}\\
 \frac{1}{n_x}\bm t_x\!\cdot(\bm x-\bm x_p)
 +t_\rho(\rho-\rho_p)&=0.
 \label{eq:app_palc_arc}
\end{align}
The rotating branch is initiated using the symmetry-breaking eigenmode at the Taylor-state instability. Two nearby converged
solutions with small nonzero rotation provide the initial continuation direction. The continuation step is adjusted automatically to maintain convergence. Accepted solutions satisfy the discrete steady equations and the concentration constraints $\Gamma\geq0$ and $\chi\Gamma<1$. Pseudo-arclength continuation allows both stable and unstable segments to be followed through folds.

Parameter sweeps use the previously converged solution as the
initial guess at the next parameter value. The rotation maps
are constructed from steady solution branches, with stability
determined by the eigenvalue analysis described below.

\subsection{Transient calculations and tangential stresses}

Transient calculations use the same spatial discretisation as
the steady calculations and the dimensionless time
$t=t_{\mathrm{dim}}/\tau^+$ used in the main text.
The histories in Fig.~9 describe relaxation in the
non-rotating Taylor regime following application of a constant
electric field at $\ReE=1$.

Time integration uses a second-order exponential Runge--Kutta
scheme (ETD2RK) \citep{sm-CoxMatthews2002}. The linear operator $L$
includes charge relaxation, diffusion, exchange and the linearised
Marangoni contribution about uniform concentration, and is diagonal
in Fourier space. Its exponential is evaluated directly, avoiding
the stability restrictions associated with explicit treatment of
these linear terms. The remaining terms are retained in
$\mathcal N$, so the splitting does not alter the nonlinear equations.

Writing the Fourier-space system as
$d\widehat{\bm y}/dt
=L\widehat{\bm y}+\mathcal N(\bm y)$, where
$\mathcal N(\bm y)$ denotes the Fourier coefficients of the
remaining terms, a step of length $\Delta t$ is
\begin{align}
 \widehat{\bm a}
 &=e^{L\Delta t}\widehat{\bm y}
 +\Delta t\,\varphi_1(L\Delta t)
 \mathcal N(\bm y),\\
 \widehat{\bm y}_{\rm next}
 &=\widehat{\bm a}
 +\Delta t\,\varphi_2(L\Delta t)
 [\mathcal N(\bm a)-\mathcal N(\bm y)],
 \label{eq:app_etd2}
\end{align}
where $\varphi_1(z)=(e^z-1)/z$ and
$\varphi_2(z)=(e^z-1-z)/z^2$, evaluated by their regular series
near zero. Electric fields, tractions and velocity are reconstructed
at each stage. One full step is compared with two half steps,
using their difference divided by three as the local error estimate.
The two-half-step state is accepted when the error and
concentration-admissibility tests pass. The time step also satisfies
an advective restriction.

The three tangential tractions plotted in the main text are
evaluated directly on the interface,
\begin{equation}
 \tau_E=qE_\theta,
 \qquad
 \tau_M=-\Ma\frac{1}{1-\chi\Gamma}
 \frac{\partial\Gamma}{\partial\theta},
 \qquad
 \tau_H=-(\tau_E+\tau_M).
 \label{eq:app_tractions}
\end{equation}
The Marangoni traction is discretised using Eq.~\eqref{eq:app_discrete_marangoni}. The hydrodynamic traction is evaluated from the Fourier Stokes operator, with multiplier $-2(1+M)|n|$ for nonzero modes and $-2$ for the mean. The final equality in Eq.~\eqref{eq:app_tractions} supplies a check of the tangential balance on resolved modes.

\subsection{Linear stability and numerical checks}

The stability routines express growth rates using the electroviscous time $\widehat t=\ReE\,t$. A computed growth
rate $\widehat\lambda$ therefore corresponds to $\lambda=\ReE\,\widehat\lambda$ in the main-text time variable. Since $\ReE>0$, this conversion leaves the stability classification unchanged.

For linear stability we linearise the semi-discrete time-dependent system represented by Eqs.~\eqref{eq:app_q_evolution} and \eqref{eq:app_G_evolution} about each steady state. For $\Ks=0$ the conserved surfactant mean defines a neutral mass mode, which is removed by restricting perturbations to zero mean; for $\Ks>0$ the corresponding mean perturbation decays at rate $\Ks$ in the time variable $t$, or $\Ks/\ReE$ in $\widehat t$, and is not neutral. The steady solve and stability calculation both use the half-turn symmetry in Eq.~\eqref{eq:app_symmetry}. The perturbations satisfy $\delta q(\theta+\pi)=-\delta q(\theta)$ and $\delta\Gamma(\theta+\pi)=\delta\Gamma(\theta)$, with both sine and cosine components included. The reported stability classifications therefore apply only to this symmetry class.

A state is classified as linearly stable to perturbations preserving the imposed half-turn symmetries when the largest real part of the computed eigenvalues, in the time variable $\widehat t$, is below $-10^{-7}$, unstable when it is above $10^{-7}$, and marginal otherwise. The onset of Quincke rotation on the Taylor branch is located from the zero crossing of the leading real eigenvalue. Along a rotating branch, changes between solid and dotted curves in the figures use the same criterion. Here bistability denotes coexistence of stable states with different rotation magnitudes over the same field interval. These may be a Taylor state and a rotating state, or two rotating states with different amplitudes. Figure~\ref{fig:sm_sfold} illustrates the latter case through a zoom of the rotating branch.

Several numerical checks are used throughout. The analytic steady Jacobian is compared with centred finite-difference directional derivatives, and the stability spectrum is checked against a direct finite-difference Jacobian for representative states. The clean-interface limit reproduces the circular-drop branches and stability changes of \citet{sm-PengSchnitzer2025}; the weak-field solution is recovered as $\ReE\to0$; and the surfactant solver approaches the clean response when Marangoni coupling is removed or when diffusion or exchange becomes sufficiently rapid.

\section{Plots with a fixed reference velocity}
\label{sec:fixed_reference}
Figures~S1--S7 reproduce Figs.~2, 3, 5, 8, 11, 12 and 13 of the main text using the fixed velocity scale
\begin{equation}
 U_{c,s}=\frac{\epsilon^+E_{c,s}^2a}{\mu^+},
 \qquad e=\frac{E_0}{E_{c,s}},
 \qquad U_E=e^2U_{c,s}.
 \label{eq:sm_Uc}
\end{equation}
Thus velocities and rotation rates are multiplied by $e^2$, and the strain intensity by $e^4$:
\begin{equation}
 \widetilde{\bu}=e^2\bu,\qquad
 \widetilde{\Omega}=e^2\Omega,\qquad
 \widetilde{\mathcal S}=e^4\mathcal S.
 \label{eq:sm_conversion}
\end{equation}
The underlying solutions, stability classifications, critical fields and flow-type parameter $\Theta$ are unchanged. Figure~6 of the main text already uses $U_{c,s}$. The straining angle in Fig.~7 is invariant under this change of velocity scale and is not repeated.

\begin{figure}[!htbp]
 \centering
 \includegraphics[width=0.9\textwidth,height=0.72\textheight,keepaspectratio]{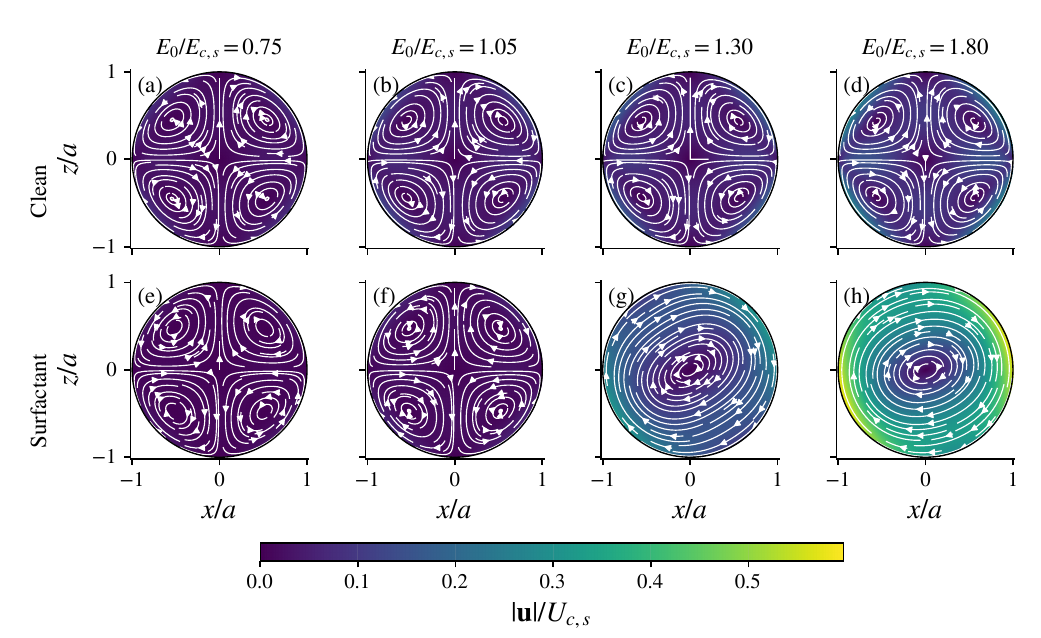}
 \caption{Same as Fig.~2 of the main text, with interior speed scaled by $U_{c,s}$. Top: clean drops. Bottom: surfactant-covered drops. All parameters and applied fields are unchanged.}
 \label{fig:sm_main2}
\end{figure}

\begin{figure}[!htbp]
 \centering
 \includegraphics[width=0.9\textwidth,height=0.72\textheight,keepaspectratio]{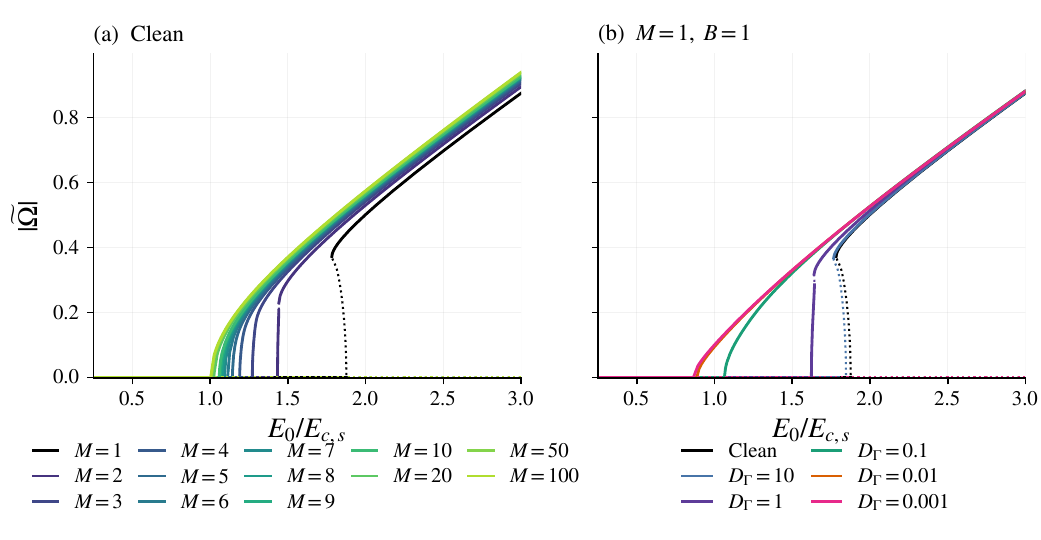}
 \caption{Same as Fig.~3 of the main text, with rotation plotted as $|\widetilde\Omega|=e^2|\Omega|$. Left: clean drops at different viscosity ratios. Right: insoluble surfactant at different surface diffusivities. The solid and dotted lines retain the main-text stability convention.}
 \label{fig:sm_main3}
\end{figure}

\begin{figure}[!htbp]
 \centering
 \includegraphics[width=0.9\textwidth,height=0.72\textheight,keepaspectratio]{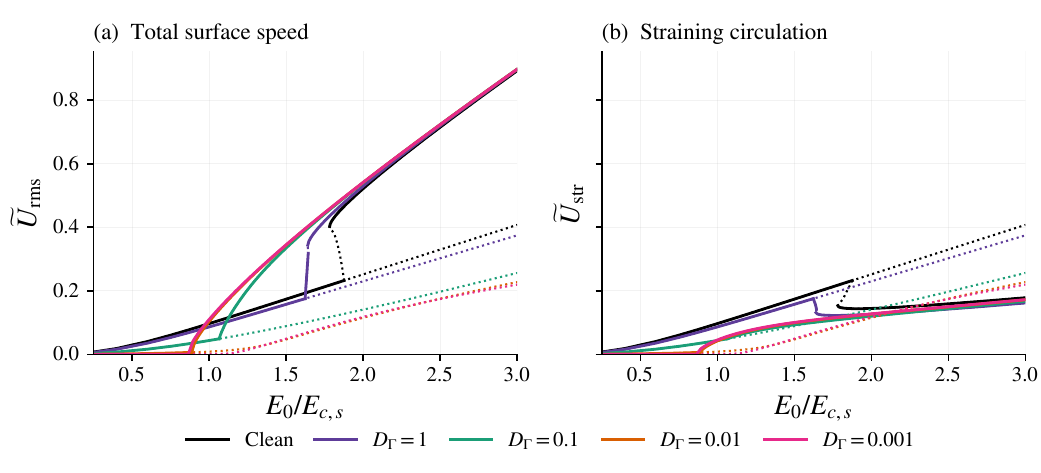}
 \caption{Same as Fig.~5 of the main text, with total and straining RMS surface speeds scaled by $U_{c,s}$: $\widetilde U_{\rm rms}=e^2U_{\rm rms}$ and $\widetilde U_{\rm str}=e^2U_{\rm str}$.}
 \label{fig:sm_main5}
\end{figure}

\begin{figure}[!htbp]
 \centering
 \includegraphics[width=0.9\textwidth,height=0.72\textheight,keepaspectratio]{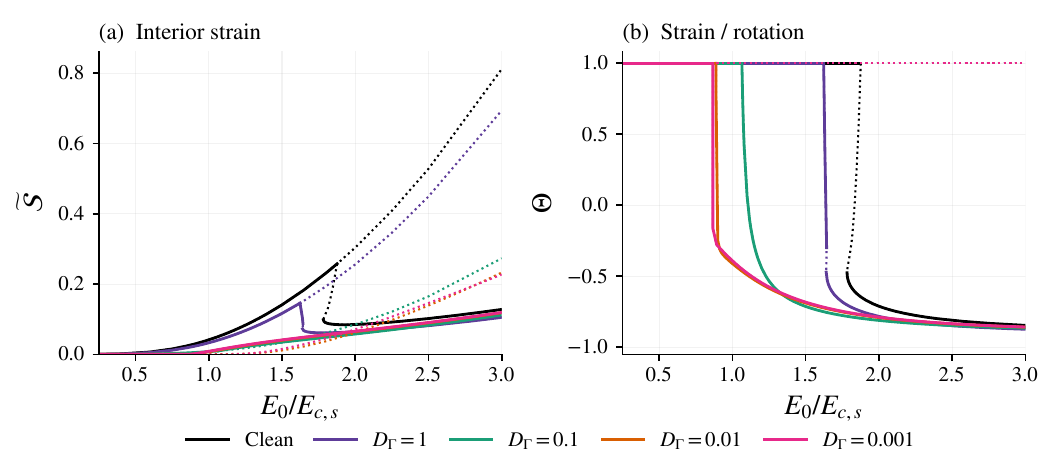}
 \caption{Same as Fig.~8 of the main text, with strain intensity plotted as $\widetilde{\mathcal S}=e^4\mathcal S$ (left). The flow-type parameter $\Theta$ (right) is unchanged.}
 \label{fig:sm_main8}
\end{figure}

\begin{figure}[!htbp]
 \centering
 \includegraphics[width=0.9\textwidth,height=0.72\textheight,keepaspectratio]{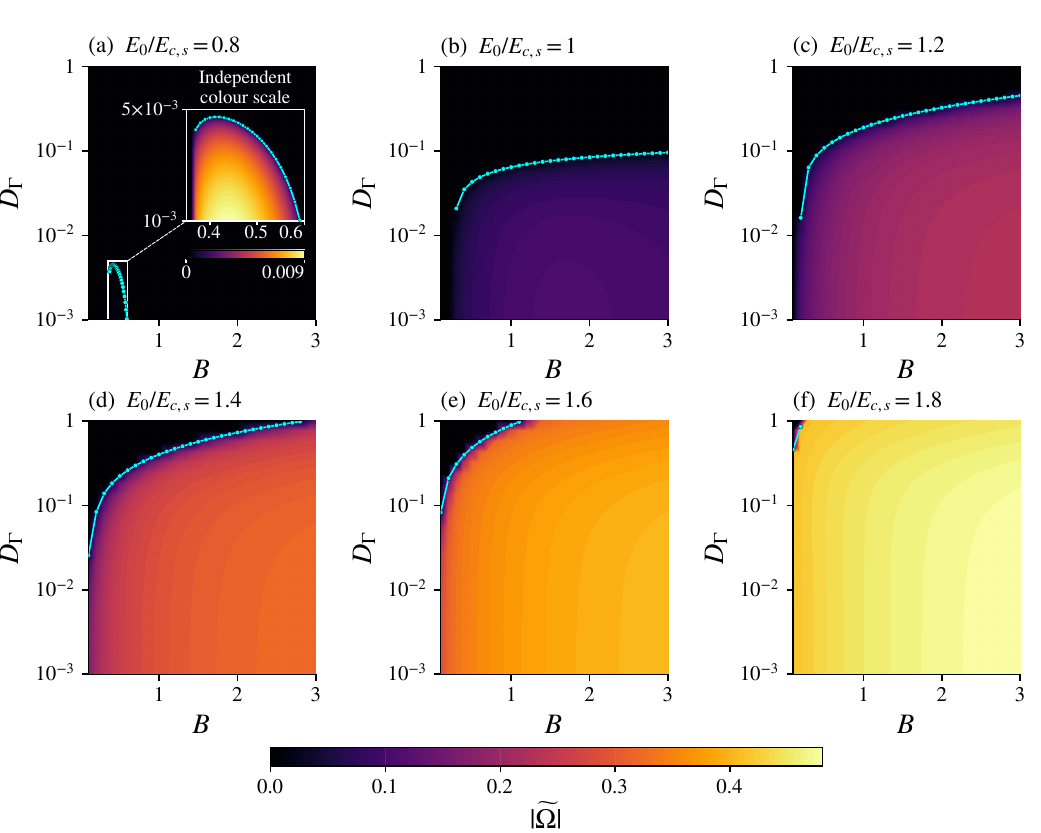}
 \caption{Same as Fig.~11 of the main text, with colour showing $|\widetilde\Omega|=e^2|\Omega|$. Each panel uses its indicated applied field. The inset retains an independent colour scale; the branch-selection and stability conventions are unchanged.}
 \label{fig:sm_main11}
\end{figure}

\begin{figure}[!htbp]
 \centering
 \includegraphics[width=0.9\textwidth,height=0.72\textheight,keepaspectratio]{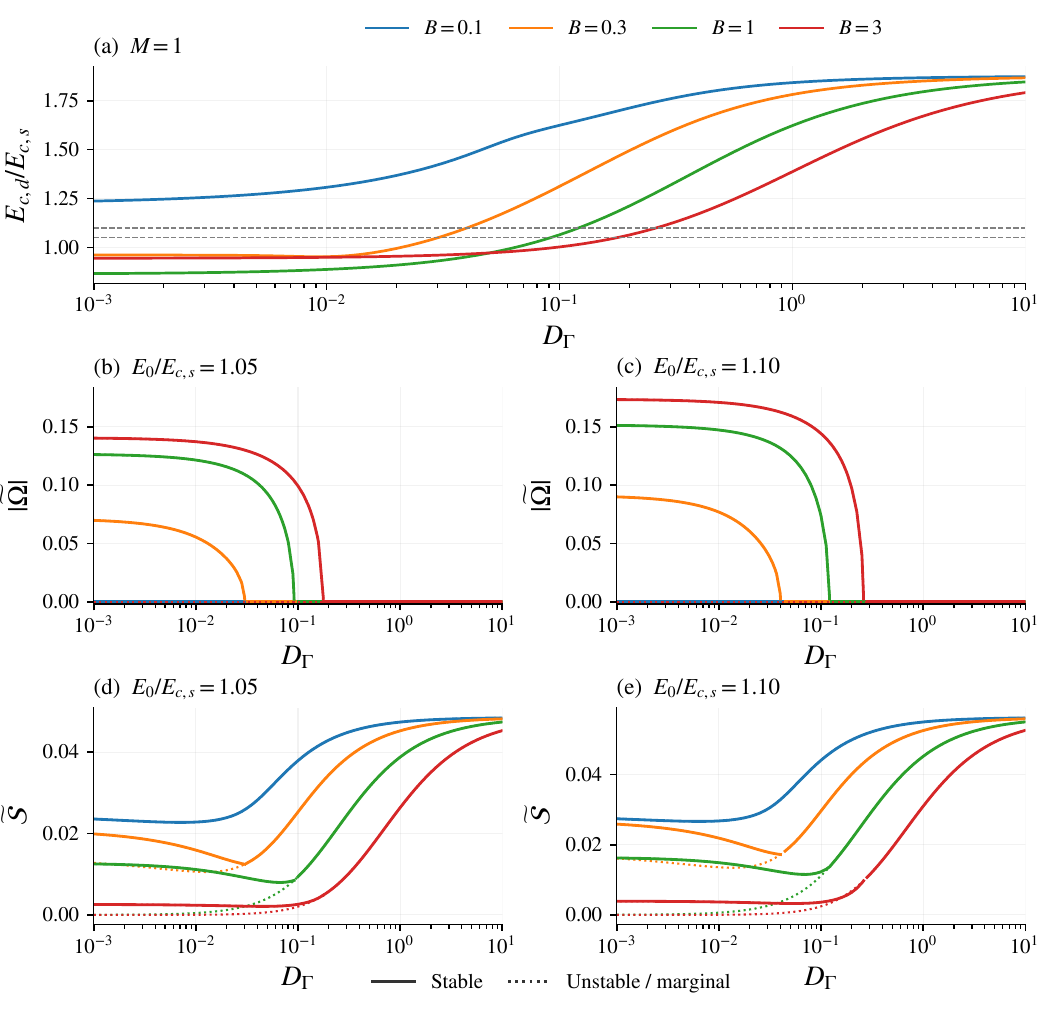}
 \caption{Same as Fig.~12 of the main text. The critical-field curves (top) are unchanged; the rotation magnitudes (middle) and strain intensities (bottom) are multiplied by $e^2$ and $e^4$, respectively, at each indicated fixed applied field.}
 \label{fig:sm_main12}
\end{figure}

\begin{figure}[!htbp]
 \centering
 \includegraphics[width=0.49\textwidth,height=0.72\textheight,keepaspectratio]{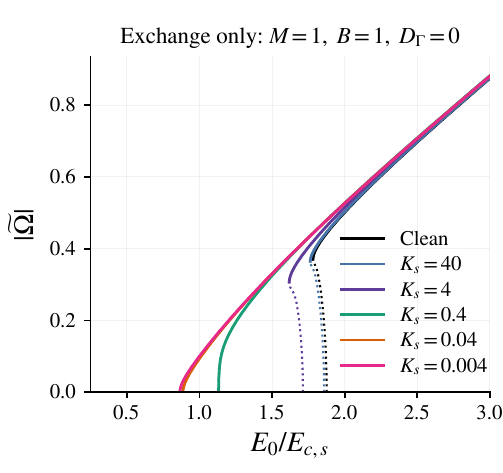}
 \caption{Same as Fig.~13 of the main text, with rotation plotted as $|\widetilde\Omega|=e^2|\Omega|$ for the exchange-only family. All parameters and stability conventions are unchanged.}
 \label{fig:sm_main13}
\end{figure}

\begin{figure}[!htbp]
 \centering
 \includegraphics[width=0.49\textwidth]{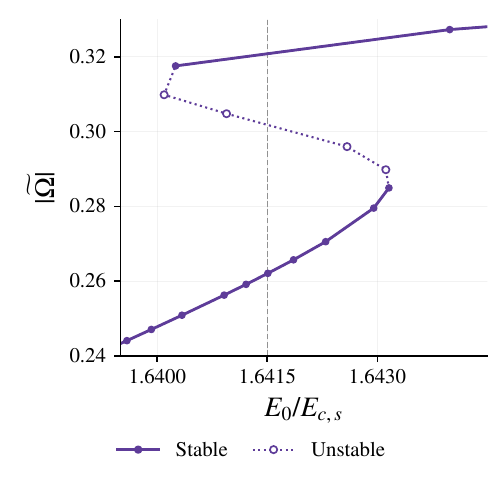}
 \caption{Finite-amplitude bistability for the insoluble-surfactant case $R=S=2$, $M=1$, $B=1$, $\DG=1$, $\chi=0.2$, $\Dq=10^{-3}$ and $\Ks=0$. Zoom of the S-shaped rotating branch in Fig.~\ref{fig:sm_main3}(b), corresponding to Fig.~3(b) of the main text, using the fixed reference scale $|\widetilde\Omega|=e^2|\Omega|$. The vertical dashed guide at $e=E_0/E_{c,s}=1.6415$ intersects two stable rotating arms with different amplitudes, separated by an unstable arm. Filled and open symbols mark saved stable and unstable solutions, respectively; solid and dotted lines join them in continuation order. The Taylor state loses stability at $e\simeq1.6229$, below the field interval shown.}
 \label{fig:sm_sfold}
\end{figure}

\section{Illustrative dimensional realisation}
To give dimensional context for the parameters used in this study, we use the castor-oil suspending-phase and silicone-oil material properties reported in the Quincke experiments of \citet{sm-SalipanteVlahovska2010} and take an illustrative drop radius $a=0.9\,\mathrm{mm}$. Specifically, we use $\epsilon^+/\epsilon_0=5.3$, $\sigma^+=4.5\times10^{-11}\,\mathrm{S\,m^{-1}}$, $\mu^+=0.69\,\mathrm{Pa\,s}$ and $T=298\,\mathrm{K}$. These values give $\tau^+\simeq1.04\,\mathrm{s}$ and $\ReE=1$ at $E_0\simeq1.19\times10^5\,\mathrm{V\,m^{-1}}=119\,\mathrm{V\,mm^{-1}}$. Table~\ref{tab:dimensional_mapping} gives one dimensional realisation of selected parameters for these specified properties; the mapping is not unique when the other dimensional quantities can vary. For example, $B=1$ corresponds to an equilibrium surface excess of about $0.24\,\mu\mathrm{mol\,m^{-2}}$, while $\DG=10^{-3}$ corresponds to a surface diffusivity of order $10^{-9}\,\mathrm{m^2\,s^{-1}}$.

\begin{table}[!htbp]
\centering
\caption{Illustrative dimensional realisation of selected dimensionless parameters for a $0.9\,\mathrm{mm}$ drop in the castor-oil suspending phase described in the text. The $\chi$ row uses the surface concentration in the $B=1$ row.}
\label{tab:dimensional_mapping}
\begin{tabular}{lll}
\hline
\shortstack[l]{dimensionless\\parameter}
& value
& \shortstack[l]{dimensional\\value} \\
\hline
$\ReE$ & $1$ & $E_0\simeq119\,\mathrm{V\,mm^{-1}}$ \\
$B$ & $1$ & $\Gamma_{\rm eq}\simeq0.24\,\mu\mathrm{mol\,m^{-2}}$ \\
$\DG$ & $10^{-3}$ & $D_\Gamma^{\mathrm{dim}}\simeq7.8\times10^{-10}\,
\mathrm{m^2\,s^{-1}}$ \\
$\Ks$ & $0.004$ & $k_{\rm ex}\simeq3.8\times10^{-3}\,\mathrm{s^{-1}}$ \\
$\chi$ & $0.2$ & $\Gamma_\infty\simeq1.2\,\mu\mathrm{mol\,m^{-2}}$ \\
\hline
\end{tabular}
\end{table}

The conversions for the transport and concentration parameters are
\begin{equation}
D_\Gamma^{\mathrm{dim}}=\frac{a^2}{\tau^+}\DG,\qquad
k_{\rm ex}=\frac{\Ks}{\tau^+},\qquad
\Gamma_{\rm eq}=\frac{\mu^+a}{\tau^+R_gT}B,\qquad
\Gamma_\infty=\frac{\Gamma_{\rm eq}}{\chi}.
\end{equation}
They apply after the remaining dimensional properties have been specified. The illustrative outer-fluid properties do not fix the independently chosen electrical ratios $R$ and $S$ of the model.

\
\section{Weak-field derivation of the surface speed in Taylor regime}
\label{sec:weak_field_reduction}
We derive the leading-order suppression of Taylor circulation due to surfactant. Surface-charge convection is weak in this limit, so the leading electric traction is unchanged by surfactant. Writing $\Gamma=1+g$, we assume $|g|\ll1$ and $|\chi g|\ll1-\chi$, while retaining finite $\chi$.The steady surfactant equation is
\begin{equation}
 \ReE\,\partial_\theta(\Gamma u_\theta)
 =\DG\,\partial_\theta^2\Gamma+\Ks(1-\Gamma).
\end{equation}
Write the leading quadrupolar fields as
\begin{equation}
 u_\theta=U_2\sin2\theta,\qquad
 \Gamma=1+G_2\cos2\theta.
\end{equation}
Linearisation about $\Gamma=1$ gives
\begin{equation}
 2\ReE U_2=-(4\DG+\Ks)G_2,
 \qquad G_2=-\frac{2\ReE}{\Lambda_2}U_2, \qquad \Lambda_2=4\DG+\Ks.
\end{equation}
The main-text comparisons use either diffusion alone or exchange alone. The linearised Langmuir traction is
\begin{equation}
 \tau_M=-\frac{\Ma}{1-\chi}\partial_\theta\Gamma
 =\frac{2\Ma G_2}{1-\chi}\sin2\theta
 =-\frac{4B}{(1-\chi)\Lambda_2}U_2\sin2\theta.
\end{equation}

The hydrodynamic coefficient follows from the Stokes solution inside and outside a circular interface. Using $u_r=r^{-1}\partial_\theta\psi$ and $u_\theta=-\partial_r\psi$, the streamfunctions that are regular inside, decay in velocity outside and satisfy $u_r(1)=0$ are
\begin{equation}
 \psi^- =\frac{U_2}{2}(r^2-r^4)\sin2\theta,
 \qquad
 \psi^+ =\frac{U_2}{2}(r^{-2}-1)\sin2\theta.
\end{equation}
Thus
\begin{equation}
 u_\theta^-=U_2(2r^3-r)\sin2\theta,
 \qquad u_\theta^+=U_2r^{-3}\sin2\theta.
\end{equation}
At $r=1$, $\partial_\theta u_r=0$ and the shear stress is proportional to $\partial_r u_\theta-u_\theta/r$. In units of $\epsilon^+E_0^2$, this gives
\begin{equation}
 T^-_{H,r\theta}=4MU_2\sin2\theta,\qquad
 T^+_{H,r\theta}=-4U_2\sin2\theta,\qquad
 \tau_H=-4(1+M)U_2\sin2\theta.
\end{equation}

Let $\tau_E=T_2\sin2\theta$ be the leading electric traction. Its amplitude is the same in the covered and clean systems at this order because charge convection enters at higher order. The balance $\tau_H+\tau_E+\tau_M=0$ gives
\begin{equation}
 T_2=\left[4(1+M)+\frac{4B}{(1-\chi)\Lambda_2}\right]U_2.
\end{equation}
For the clean interface, $T_2=4(1+M)U_{2,\mathrm{clean}}$. Eliminating $T_2$ therefore yields
\begin{equation}
 \frac{U_2}{U_{2,\mathrm{clean}}}
 =\left[1+\frac{B}{(1+M)(1-\chi)\Lambda_2}\right]^{-1}.
 \label{eq:sm_weak_speed}
\end{equation}
For the quadrupolar field, $U_{\rm str}=|U_2|/\sqrt{2}$, so the same ratio applies to the RMS straining surface speed. Matching $\Lambda_2$ therefore gives the same leading-order surface-speed ratio for diffusion and exchange. At finite electric field, changes in charge convection alter the electric traction, and higher harmonics of velocity and concentration contribute. Equation~\eqref{eq:sm_weak_speed} is compared with the full numerical results in Fig.~14(a) of the main text.

\section{Supplementary movies}
\label{sec:movies}

Movies~S1--S6 show the evolution of surface charge, surfactant concentration
where present, and the interior flow following application of a constant
electric field. The common parameters are $R=S=2$, $M=1$ and
$\Dq=10^{-3}$. The surfactant-covered cases have $B=1$ and $\chi=0.2$;
the clean cases have no surfactant and $B=0$. The applied field, transport
parameters and number of uniform angular grid points are listed in
Table~\ref{tab:movie_parameters}. The field reference is the same rigid-cylinder
value as in the main text, with $\mathrm{Re}_{E,c,s}=1.5060$ and
$\ReE=1.5060(E_0/E_{c,s})^2$.

\begin{table}[!htbp]
\centering
\caption{Parameters of the supplementary movies. The regime is the final
state shown. A dash indicates a surfactant parameter that does not apply
to the clean interface.}
\label{tab:movie_parameters}
\small
\setlength{\tabcolsep}{7pt}
\begin{tabular}{cllcccc}
\hline
Movie & Interface & Regime & $E_0/E_{c,s}$ & $\DG$ & $\Ks$ & $N$ \\
\hline
S1 & Clean & Taylor & $1.50$ & -- & -- & $2048$ \\
S2 & Clean & Quincke & $2.00$ & -- & -- & $2048$ \\
S3 & Insoluble surfactant & Taylor & $0.85$ & $10^{-3}$ & $0$ & $1024$ \\
S4 & Insoluble surfactant & Quincke & $1.00$ & $10^{-3}$ & $0$ & $1024$ \\
S5 & Soluble surfactant & Taylor & $0.85$ & $0$ & $4$ & $2048$ \\
S6 & Soluble surfactant & Quincke & $2.00$ & $0$ & $4$ & $4096$ \\
\hline
\end{tabular}
\end{table}

All movies begin with an uncharged interface, $q(\theta,0)=0$.
For the surfactant-covered cases, $\Gamma(\theta,0)=1$.
Movies~S1, S3 and S5 follow relaxation to a non-rotating Taylor state.
In Movies~S2 and S6, a small perturbation
$\delta q=-10^{-3}\sin\theta$ is applied at $t=6$ to initiate the
rotational instability. Movie~S4 first approaches the Taylor state while
reflection symmetry is imposed. This reflection constraint is then removed
and a perturbation $\delta q=-10^{-4}\sin\theta$ initiates rotation.
The applied electric field and all material parameters remain unchanged
during each movie, including at the perturbation. The perturbations preserve
zero net charge and do not change the surfactant concentration.

The displayed time is $t=t_{\rm dim}/\tau^+$, labelled $T$ in the movies.
Surface speeds are scaled by $U_E$, and angular
velocities by $U_E/a$. The interior velocity is reconstructed from the
interfacial solution at each time. Interior tracers follow this velocity.
The charge symbols represent weighted surface-charge patches whose weights
change through Ohmic currents and diffusive transport. For the insoluble
surfactant, surface markers represent the combined advective and diffusive
surfactant flux. In the exchange movies, their weights also change through
the local source $\Ks(1-\Gamma)$.

Movies~S3 and S5 compare weak surface diffusion with rapid exchange at the
same field. Movies~S2 and S6 compare clean and rapidly exchanging interfaces
at the same field. These movie pairs illustrate the different transport
regimes; the matched-relaxation comparisons are those in Figs.~14--16 of
the main text.

\end{document}